\RequirePackage{rotating}
\documentclass[fleqn,usenatbib]{mnras}

\usepackage{newtxtext,newtxmath}
\usepackage[T1]{fontenc}
\usepackage{graphicx}
\usepackage{amsmath}    
\usepackage{longtable}
\usepackage{verbatim}
\usepackage[singlelinecheck=off]{caption}
\usepackage{pdflscape}
\usepackage{booktabs, threeparttable}
\usepackage{multirow}
\usepackage{lineno}
\newcommand{\tabitem}{~~\llap{\textbullet}~~}
\usepackage{lscape}
\newcommand{\angstrom}{\text{\normalfont\AA}}

\usepackage{eso-pic}

\AddToShipoutPictureBG*{%
  \AtPageUpperLeft{%
    \hspace{0.75\paperwidth}%
    \raisebox{-3.5\baselineskip}{%
      \makebox[0pt][l]{\textnormal{DES-2022-0726}}
}}}%

\AddToShipoutPictureBG*{%
  \AtPageUpperLeft{%
    \hspace{0.75\paperwidth}%
    \raisebox{-4.5\baselineskip}{%
      \makebox[0pt][l]{\textnormal{FERMILAB-PUB-22-837-PPD}}
}}}%

\title[Ultracool dwarfs candidates from DES Y6]{Ultracool dwarfs candidates based on six years of the Dark Energy Survey data}

\author[DES Collaboration]{
\parbox{\textwidth}{
\Large
M.~dal Ponte,$^{1,2}$
B.~Santiago,$^{1,2}$
A.~Carnero~Rosell,$^{3,2,4}$
L.~De Paris,$^{1}$
A.~B.~Pace,$^{5}$
K.~Bechtol,$^{6}$
T.~M.~C.~Abbott,$^{7}$
M.~Aguena,$^{2}$
S.~Allam,$^{8}$
O.~Alves,$^{9}$
D.~Bacon,$^{10}$
E.~Bertin,$^{11,12}$
S.~Bocquet,$^{13}$
D.~Brooks,$^{14}$
D.~L.~Burke,$^{15,16}$
M.~Carrasco~Kind,$^{17,18}$
J.~Carretero,$^{19}$
C.~Conselice,$^{20,21}$
M.~Costanzi,$^{22,23,24}$
S.~Desai,$^{25}$
J.~De~Vicente,$^{26}$
P.~Doel,$^{14}$
S.~Everett,$^{27}$
I.~Ferrero,$^{28}$
B.~Flaugher,$^{8}$
J.~Frieman,$^{8,29}$
J.~Garc\'ia-Bellido,$^{30}$
D.~W.~Gerdes,$^{31,9}$
R.~A.~Gruendl,$^{17,18}$
D.~Gruen,$^{13}$
G.~Gutierrez,$^{8}$
S.~R.~Hinton,$^{32}$
D.~L.~Hollowood,$^{33}$
D.~J.~James,$^{34}$
K.~Kuehn,$^{35,36}$
N.~Kuropatkin,$^{8}$
J.~L.~Marshall,$^{38}$
J. Mena-Fern{\'a}ndez,$^{26}$
F.~Menanteau,$^{17,18}$
R.~Miquel,$^{38,19}$
R.~L.~C.~Ogando,$^{39}$
A.~Palmese,$^{40}$
F.~Paz-Chinch\'{o}n,$^{17,41}$
M.~E.~S.~Pereira,$^{42}$
A.~A.~Plazas~Malag\'on,$^{43}$
A.~Pieres,$^{2,39}$
M.~Raveri,$^{44}$
M.~Rodriguez-Monroy,$^{26}$
E.~Sanchez,$^{26}$
V.~Scarpine,$^{8}$
M.~Schubnell,$^{9}$
I.~Sevilla-Noarbe,$^{26}$
M.~Smith,$^{45}$
M.~Soares-Santos,$^{9}$
E.~Suchyta,$^{46}$
M.~E.~C.~Swanson,$^{19}$
G.~Tarle,$^{9}$
D.~Thomas,$^{10}$
C.~To,$^{47}$
N.~Weaverdyck$^{9,48}$
\begin{center} (DES Collaboration) \end{center}
}
\\
\\
Affiliations are listed at the end of the paper.
}

\date{Accepted XXX. Received YYY; in original form ZZZ}
\pubyear{2022}

\begin{document}
\label{firstpage}

\maketitle

\begin{abstract}
We present a sample of 19,583 ultracool dwarf candidates brighter than z $\leq 23$ selected from the Dark Energy Survey DR2 coadd data matched to VHS DR6, VIKING DR5 and AllWISE covering $\sim$ 4,800 $deg^2$. The ultracool candidates were first pre-selected based on their (i-z), (z-Y), and (Y-J) colours. They were further classified using a method that compares their optical, near-infrared and mid-infrared colours against templates of M, L and T dwarfs. 14,099 objects are presented as new L and T candidates and the remaining objects are from the literature, including 5,342 candidates from our previous work. Using this new and deeper sample of ultracool dwarf candidates we also present: 20 new candidate members to nearby young moving groups (YMG) and associations, variable candidate sources and four new wide binary systems composed of two ultracool dwarfs. Finally, we also show the spectra of twelve new ultracool dwarfs discovered by our group and presented here for the first time. These spectroscopically confirmed objects are a sanity check of our selection of ultracool dwarfs and photometric classification method.

\end{abstract}


\section{Introduction}

Ultracool dwarfs (UCDs) are very cool (T$_{\rm eff}$ < 2700 K), low mass (M < 0.1 M${_\odot}$) objects, ranging from spectral type M7 and later. They include both very low mass stars and brown dwarfs. Brown dwarfs are not massive enough ($\sim$ 0.072 M$_{\odot}$) to burn hydrogen in their core. Therefore, they continue to cool and dim over time across spectral types M, L, T and Y \citep{Kirkpatrick1999, Cushing2011}. Without sustained hydrogen fusion, there is a degeneracy between mass, age and luminosity. Their spectra are characterized by the effects of clouds and molecular absorption bands. For the L dwarfs, the spectra in the red optical is characterized by the weakening of TiO and VO, strengthening of FeH, CrH, H$_{2}$O, and alkali metals such Na I, K I, Cs I, Rb I \citep{Kirkpatrick1999}. The mid infrared spectra are similar to M dwarfs, with H$_{2}$O and CO as the most prominent bands along with the presence of clouds in the photosphere \citep{Burgasser2002a}. The transition sequence to the T dwarfs is characterized by the disappearance of clouds from the photosphere, leading to relatively bluer colours in the near-infrared (NIR) compared to the L sequence. Their spectra is characterized by strong absorption features of H$_{2}$O, CH$_{4}$, and CIA H$_{2}$ \citep{Burgasser2002b}.

Despite UCDs being a very common type of object in the Galaxy, roughly 1/6 of the local stellar population by number density, they are very difficult to detect at larger distances due to their faint luminosities. Large samples of UCDs from wide-field imaging surveys (eg. Two-Micron All-Sky Survey \citep[2MASS; ][]{Skrutskie2006}, Deep Near Infrared Survey of the Southern Sky \citep[DENIS; ][]{Epchtein1997}, UKIRT Infrared Deep Sky Survey \citep[UKIDSS; ][]{lawrence07}, Wide-field Infrared Survey Explorer \citep[WISE;][]{Wright2010}, VISTA Hemisphere Survey \citep[VHS; ][]{mcmahon13}) have been discovered and revealed many important features about the ultracool dwarfs population. However, the census is still heterogeneous and shallow. The accurate identification and classification of ultracool dwarfs in wide deep ground-based surveys using only photometry enables the creation of homogeneous samples without relying on extensive spectroscopic campaigns. These samples are essential to measuring the luminosity and mass functions \citep{Cruz2007, Bochanski2010} of the ultracool dwarfs in the Galaxy, the disk scale height \citep{Ryan2005, Sorahana2019, carnero2019}, the frequency of close and wide binaries \citep{Luhman2012, Dhital2015, Fontanive2018} and the kinematics \citep{Faherty2010, Faherty2012, Smith2014, Best2018}.

Taking advantage of the Dark Energy Survey (DES) depth in the optical bands $i$, $z$ and $Y$, it is possible to select a large homogeneous sample of UCD candidates to greater distances. \citet{carnero2019} were able to select a sample of 11,745 L and T dwarf candidates using the first three years of the DES along with VHS and AllWISE \citep{Cutri2013} data. Here we expand the search for ultracool dwarfs candidates using the full six years of DES observations. Comparing to \citet{carnero2019}, the DES data are now photometrically deeper, with more reliable/precise photometry. This will allow us to probe fainter candidates, increasing the previous samples of L and T dwarfs. Besides, we have now available a sky coverage of almost the entire DES footprint, whereas in \citet{carnero2019} we only had $\sim$ 2,400 $deg^2$. This is due to the new data releases of VHS and VIKING surveys that are also used in the analyses. 

The paper outline is as follows. In Section \ref{sec:data} we present the photometric data used in this work. In Section \ref{sec:templates_colour} we present the updated colour templates for M dwarfs and UCDs and the colour cuts used to pre-select our candidates. In section \ref{sec:classification} we discuss the photometric classification methodology, where we estimate a spectral type for each target using only their photometry. In Section \ref{sec:sample_characteristics} we compare our photo-type to those of known candidates from the literature and discuss the contamination by extragalactic sources. In Section \ref{sec:sample_use}, we present several uses for our L and T dwarf candidates: i) new young moving group and association candidate members; ii) photometric variable sources; iii) new binary systems constituted by two ultracool dwarfs. In Section \ref{sec:spectra} we show the spectra of twelve new ultracool dwarfs presented previously in the \citet{carnero2019} catalog that supports our selection of ultracool dwarfs and photometric classification method. Finally, in Section \ref{sec:conclusions}, we present our conclusions.

\section{Data}
\label{sec:data}
\subsection{DES, VHS, VIKING and AllWISE}

DES is a $\sim$5,000 $deg^2$ optical survey in the $grizY$ bands that used the Dark Energy Camera \citep[DECam;][]{flaugher2015}. DECam is a wide-field (3 $\deg^2$) imager at the prime focus of the Blanco 4m telescope at Cerro Tololo Inter-American Observatory (CTIO). DES observations started in September 2013 and were completed in January 2019, spanning nearly six years. 

DES DR2 is the assembled dataset from 6 years of DES science operations, with data collected over 681 nights and which includes ~691 million astronomical objects detected in 10,169 coadded image tiles of size 0.534 $\deg^2$ produced from 76,217 single-epoch images. The estimated area loss to image defects, saturated stars, satellite trails, etc is of $\simeq$ 200 $deg^2$. After a basic quality selection, galaxy and stellar samples contain 543 million and 145 million objects, respectively. The typical depths (in AB system) as estimated from the magnitude at S/N = 10 in the coadd images are $g$ = 24.0, $r$ = 23.8, $i$ = 23.1, $z$ = 22.13, and $Y$ = 20.7 \citep{abbott2021}.

For the purpose of our work, we matched the DES DR2 catalog to the VHS DR6, VIKING DR5 \citep{Edge2013} and AllWISE catalogs using a positional matching radius of $2\arcsec$, keeping only the best match, i.e, the nearest object found.
The DES+VHS coverage area is around 4,500 $deg^2$. The VHS survey is imaged with exposure time per coadded image of 120–240 seconds in $J$ and 120 seconds in $K_{s}$. There is also partial coverage in the $H$ band with an exposure time of 120 seconds. The median 5$\sigma$ point source depths is $J_{AB} \sim$ 21.4, $H_{AB} \sim$ 20.7 and $K_{s,AB} \sim$ 20.3. Since, by design, the VIKING and VHS footprints are complementary, we decided to use also the VIKING DR5 data in regions not covered by VHS. The DES+VIKING coverage is about 500 $deg^2$, providing along with VHS, almost the entire DES footprint. VIKING has a median depths at 5$\sigma$ of $J_{AB} \sim$ 22.1, $H_{AB} \sim$ 21.5 and $K_{s,AB} \sim$ 21.2 across all imaged regions ($\sim$ 1350 $deg^2$). 
Lastly, for the AllWISE survey we will use only $W1$ and $W2$ bands, which is $>$95\% complete for sources with $W1 < 17.1$ and $W2 < 15.7$ (in Vega system).

Some quality cuts were initially applied to the matched catalog, such as \verb+IMAGFLAGS_ISO_i,z,Y+ = 0 from DES DR2 and \verb+J,H,K_{s}ppErrBits+ < 255, to ensure that the object has not been affected by spurious events in the images in $i$, $z$, $Y$, $J$, $H$ and $K_{s}$ bands. We also imposed a magnitude limit cut of $z$ < 23 (DES) and a simultaneous 5$\sigma$ level detection in the $i$, $z$, $Y$ (DES) and $J$ (VHS+VIKING). We did not apply any standard star/galaxy separation because they are not as efficient for relatively nearby sources with significant signature of proper motions on their coadded DES images. In this work, we adopted the \verb+PSF_MAG_i,z,Y+ magnitude type from DES and \verb+apermag3_J,H,Ks+ from VHS and VIKING catalogs. Also, all DES magnitudes and colours are in the AB system and the VHS+VIKING and AllWISE magnitudes and colours are in the Vega system. 

It is important to mention that for sources with significant proper motions, a matching radius of 2 $\arcsec$ may be too small. This matching radius will work except for the very nearby (<= 6pc) or high-velocity (> 50 km/s) cases. Therefore, a small percentage of ultracool dwarfs will be missing from our catalogue due to this effect. The matching between DES data and others surveys provides a broad photometric baseline, spanning from the optical to the infrared. All these bands will be later used to construct empirical templates, perform the colour selection and photometrically estimate the spectral type of our UCDs candidates. The entire selection and classification process is summarized in Table \ref{tab:colorcuts} where every step is highlighted along with the corresponding section in this paper.

\begin{table}
\centering
\caption{Steps used in this paper to select and classify L and T dwarfs using DES+VHS+VIKING+AllWISE. First, a magnitude limit is imposed in the $z$ band, quality cuts are applied to the data to remove spurious targets and colour cuts ($i-z$), ($z-Y$) and ($Y-J$) are applied to select only the reddest objects. These are the sources that enter into the classification method. Next, we imposed that every object must have six or more bands and spectral type L0 or later. Then, extragalactic contamination is removed and the proper motion is assessed to recover objects erroneously assigned as extragalactic sources. Finally, we list candidates previously found in the literature and new ones.}
\label{tab:colorcuts}
\scalebox{0.8}{
\begin{tabular}{cccc}
\hline 
Step & Description & Number of Targets & Section \\ 
\hline \hline
0    & DES Y6 (DR2) & 691,483,608 & 2.1\\
\hline
\multirow{5}*{1} 
     & $z<23$ & & \\
     & SNR$_{z,Y}$ > 5$\sigma$ & &  \\
     & IMAFLAGS\_ISO\_i,z,Y=0 & 602,366 &  2.1\\
     & $(i-z)_{AB} > 1.20$ & & \\
     & $(z-Y)_{AB} > 0.15$ & & \\
\hline
\multirow{3}*{2}
     & Matching 2$\arcsec$ DES +VHS+VIKING & & \\
     & $(Y_{AB}-J_{Vega}) > 1.55$ & 164,406 & 2.1\\ 
     & SNR$_{J}$ > 5$\sigma$ & & \\
     & ${J,H,K_{s}}$ppErrBits > 256 & & \\
\hline
3    & Matching 2$\arcsec$ DES + AllWISE & 76,184 & 3 \\
\hline
4    & Photo-Type classification $\geq$ L0 & 53,565 & 4 \\
\hline
5    & After removal of extragalactic contamination & 19,449 & 5.2 \\  
\hline
6    & Recover by proper motion criterion & 141 & 5.3 \\
\hline
7    & From the literature & 5,484 & 5.4 \\
\hline
8    & New candidates & 14,099 & 5.5 \\
\hline
\end{tabular}
}
\end{table}

\subsection{Known ultracool dwarfs}

The sample from \cite{best2020} (hereafter B2020 sample) contains the most up-to-date compilation of ultracool dwarfs with spectroscopic confirmation. The complete sample has 2,940 sources, with spectral type ranging from M3 to Y2. This compilation includes spectral types from optical and NIR. When both are available for a source, the authors recommend using optical types for M and L dwarfs and NIR types for T dwarfs, given that these are the spectral domains of the dominant features required for spectral classification in each case.
From this catalogue there are 388 sources located in the DES footprint, and 292 of them are classified as L or T dwarfs. For the construction of the templates, we excluded objects flagged in the B2020 sample as unresolved binaries and sub-dwarfs. We first matched the B2020 sample of L and T dwarfs with the DES DR2 catalog and found 227 objects in common. Since we have a small number of objects between the B2020 sample and DES, we decided to adopt only in this step a positional match of 3$\arcsec$. Every matched source was inspected visually using the DES image portal tool. The remaining 65 objects were eliminated in our selection due to quality cuts or for having a positional match beyond the limit. Then we matched B2020+DES DR2 with VHS DR6 and VIKING DR5 resulting in 185 objects in common. The 42 lost objects in the match between DES and VHS+VIKING are due to lack of data or positional match beyond 2$\arcsec$ or the VHS+VIKING quality flag applied. Finally, we matched all the B2020 sample with a combination of VHS+VIKING+AllWISE, regardless of DES data, and we end up with 658 objects. We take these three steps in order to obtain as many objects as possible to construct our colour templates. In comparison with the sample of known ultracool dwarfs in \citet{carnero2019}, there are more 19 objects with DES magnitudes, 81 more in DES+VHS+VIKING and 530 more in VHS+VIKING+AllWISE. Here, the difference between the samples with and without DES data is due to the limited area of the south where DES footprint is located. The combination of VHS+VIKING+AllWISE covers almost the entire southern hemisphere.

\subsection{Known contaminants}

There are two main types of sources that we consider as contaminants at this stage: M dwarfs and quasars at high redshift. In \citet{carnero2019} we used a sample of 70,841 visually inspected M dwarfs from \citet{west11}. Here we use the \citet{Kiman2019} compilation of spectroscopic confirmed 73,473 M and 743 L or later dwarfs from SDSS constructed from \citet{west11}, \citet{Schmidt2015} and \citet{Schmidt2019}. The match between \citet{Kiman2019} and DES DR2, VHS DR6 and AllWISE data resulted in 19,355 objects in common. This updated M dwarfs sample, with new DES photometry, is fundamental for the update of our colour templates, used in the classification scheme. Regarding the quasars, we are now using the quasar catalogue from SDSS DR16 presented by \citet{Lyke2020}. For this latter, we only kept objects with redshift $z> 4$. The reason is that the low-z quasars have much bluer colors than the UCDs and therefore are not relevant to our contamination analysis.

\section{Templates and colour selection}
\label{sec:templates_colour}

We updated our empirical colour templates using the samples of known M, L and T dwarfs described previously. 
The construction of the templates followed the same methodology described in \citet{carnero2019}. For the M dwarfs (M0 to M9), we used the median color for each spectral type as the template value. We demanded SNR $>$ 5$\sigma$ in all bands and excluded objects that were $> 2~\sigma$ from the median. The median was then recalculated after these outliers were removed in an interative process until convergence. For the L and T dwarfs, because of the smaller number of objects, we fit a $n$ order polynomial to each colour vs. spectral type relation, using the least squares method. For ($i$-$z$), ($J$-$K_{s}$), ($H$-$K_{s}$) and ($K_{s}$-$W1$) an order 4 polynomial was used; ($Y$-$J$) and ($W1$-$W2$) an order 3 and ($z$-$Y$) order 2 polynomial were used.

\begin{table}
\centering
\caption{Updated template colours of M0–T9 dwarfs.}
\label{tab:templates}
\scalebox{0.78}{
\begin{tabular}{llllllll}
\hline\hline
SpT & i-z & z-Y & Y-J & J-H & H-Ks & Ks-W1 & W1-W2 \\
\hline
M0 & 0.28 & 0.08 & 1.12 & 0.59  & 0.17 & 0.09 & 0.01 \\
M1 & 0.35 & 0.10 & 1.14 & 0.57  & 0.20 & 0.12 & 0.05 \\
M2 & 0.42 & 0.12 & 1.17 & 0.55  & 0.22 & 0.13 & 0.09 \\
M3 & 0.50 & 0.14 & 1.20 & 0.53  & 0.23 & 0.15 & 0.13 \\
M4 & 0.58 & 0.16 & 1.23 & 0.52  & 0.25 & 0.17 & 0.15 \\
M5 & 0.67 & 0.19 & 1.27 & 0.51  & 0.27 & 0.18 & 0.18 \\
M6 & 0.81 & 0.24 & 1.34 & 0.51  & 0.30 & 0.20 & 0.19 \\
M7 & 0.98 & 0.30 & 1.42 & 0.52  & 0.34 & 0.22 & 0.20 \\
M8 & 1.18 & 0.37 & 1.53 & 0.54  & 0.37 & 0.23 & 0.19 \\
M9 & 1.37 & 0.44 & 1.63 & 0.57  & 0.42 & 0.26 & 0.23 \\
L0 & 1.53 & 0.55 & 1.92 & 0.63  & 0.49 & 0.40 & 0.32 \\
L1 & 1.53 & 0.54 & 2.05 & 0.63  & 0.52 & 0.41 & 0.31 \\
L2 & 1.54 & 0.54 & 2.15 & 0.68  & 0.56 & 0.47 & 0.31 \\
L3 & 1.56 & 0.55 & 2.23 & 0.76  & 0.60 & 0.56 & 0.32 \\
L4 & 1.61 & 0.56 & 2.27 & 0.84  & 0.64 & 0.66 & 0.33 \\
L5 & 1.68 & 0.58 & 2.30 & 0.92  & 0.66 & 0.74 & 0.35 \\
L6 & 1.78 & 0.60 & 2.32 & 0.97  & 0.67 & 0.81 & 0.38 \\
L7 & 1.92 & 0.63 & 2.32 & 0.99  & 0.65 & 0.85 & 0.43 \\
L8 & 2.08 & 0.66 & 2.31 & 0.97  & 0.62 & 0.86 & 0.49 \\
L9 & 2.26 & 0.69 & 2.30 & 0.91  & 0.57 & 0.83 & 0.57 \\
T0 & 2.46 & 0.74 & 2.29 & 0.80  & 0.50 & 0.78 & 0.68 \\
T1 & 2.68 & 0.78 & 2.29 & 0.66  & 0.42 & 0.70 & 0.81 \\
T2 & 2.89 & 0.84 & 2.30 & 0.49  & 0.33 & 0.60 & 0.96 \\
T3 & 3.09 & 0.90 & 2.32 & 0.30  & 0.24 & 0.51 & 1.15 \\
T4 & 3.26 & 0.96 & 2.36 & 0.09  & 0.15 & 0.42 & 1.36 \\
T5 & 3.39 & 1.03 & 2.42 & -0.09 & 0.07 & 0.38 & 1.61 \\
T6 & 3.46 & 1.10 & 2.51 & -0.25 & 0.02 & 0.40 & 1.90 \\
T7 & 3.45 & 1.18 & 2.62 & -0.36 & 0.01 & 0.50 & 2.22 \\
T8 & 3.33 & 1.26 & 2.78 & -0.39 & 0.04 & 0.72 & 2.59 \\
T9 & 3.08 & 1.35 & 2.97 & -0.30 & 0.15 & 1.10 & 3.00 \\
\hline
\end{tabular}
}
\end{table}

\begin{figure*}
\begin{center}
\includegraphics[width=0.7\linewidth]{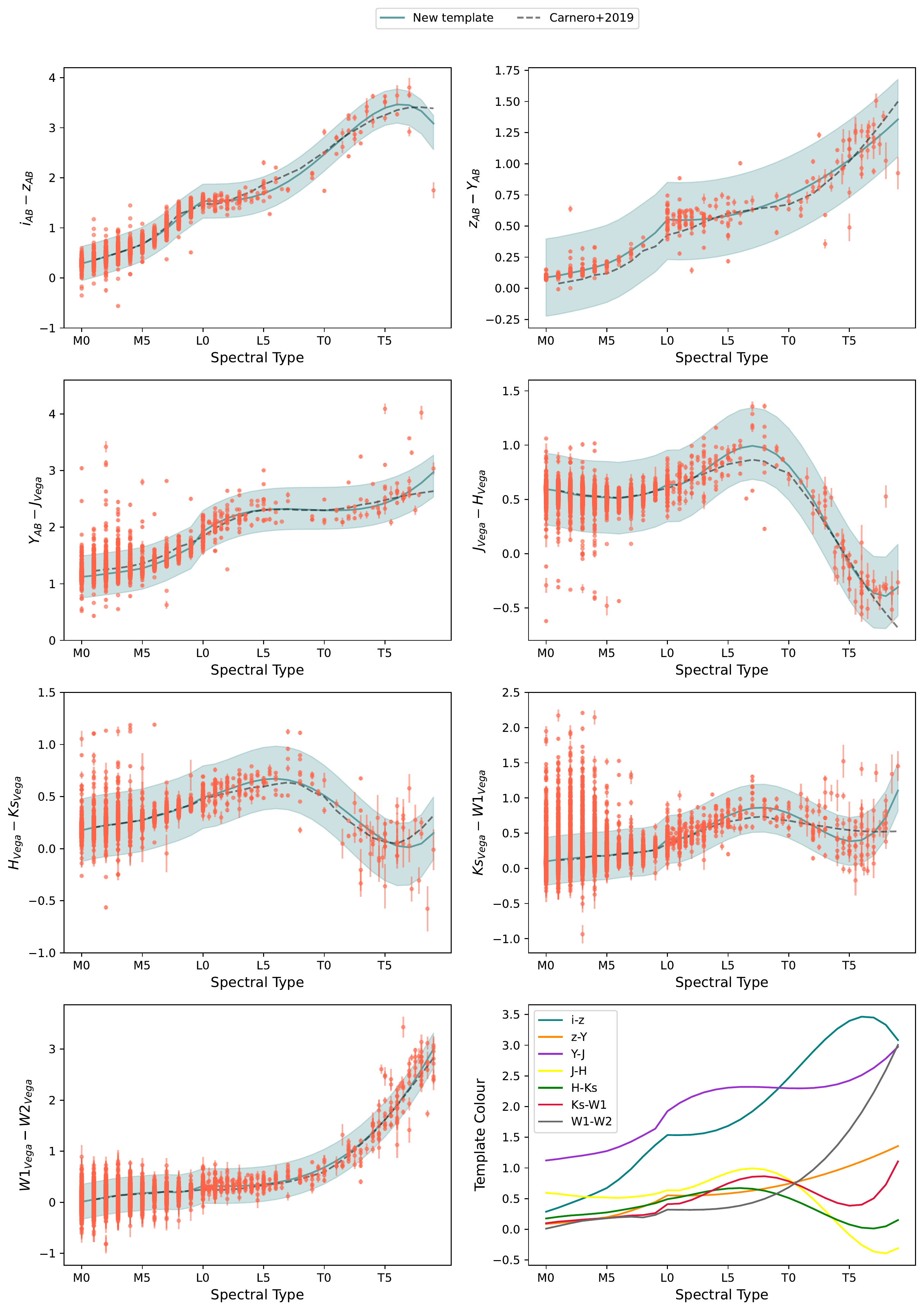}
\end{center}
\caption{Colours as a function of the spectral type for the enlarged sample of known UCDs as described in the text. The dashed line indicates the new templates, as discussed previously in Section \ref{sec:templates_colour}, and the solid line refers to the templates presented in \citet{carnero2019}. The light-shaded area corresponds to the intrinsic scatter of each colour. The last panel shows all the new updated templates for each colour indices used in this work.}
\label{fig:templates}
\end{figure*}

We re-estimated the intrinsic scatter for each colour index, assuming it to be the same for all spectral types. This intrinsic scatter is the spread in colour due to variations in metallicity, surface gravity, cloud cover, and also the uncertainty in the spectral classification. The procedure to estimate this intrinsic scatter followed the \citet{skrzypek14} prescription. We initially adopted a first guess of intrinsic scatter as 0.5 mag and added it in quadrature to the photometric errors to all templates. This new uncertainty was used to weight the points in the polynomial regression to the colour vs. spectral type relation. Then, we re-estimated the intrinsic scatter as the variance of the best-fit residuals with the rms value of the photometric errors subtracted in quadrature from it. This new value was taken as our intrinsic scatter for that colour index, irrespective of spectral type. Finally, we re-fitted the polynomial for L and T dwarfs, using the new intrinsic scatter. The intrinsic scatter values found with this method are the following: $\sigma_{i-z}=0.34, \sigma_{z-Y}=0.30, \sigma_{Y-J}=0.37, \sigma_{J-H}=0.32, \sigma_{H-Ks}= 0.30, \sigma_{Ks-W1}=0.33, \sigma_{W1-W2}=0.34$. These values are slightly smaller than those presented by \citet{dupuy12} but more aligned with those presented recently in \citet{Kirkpatrick2021b}. Even though there might be a systematic increase with spectral type, we will adopt a single value of 0.2 mag for each magnitude, corresponding to 0.3 mag for each colour index. These will later to be used to perform the spectral classification of our target sample.

\begin{figure*}
\begin{center}
    \includegraphics[width=0.8\linewidth]{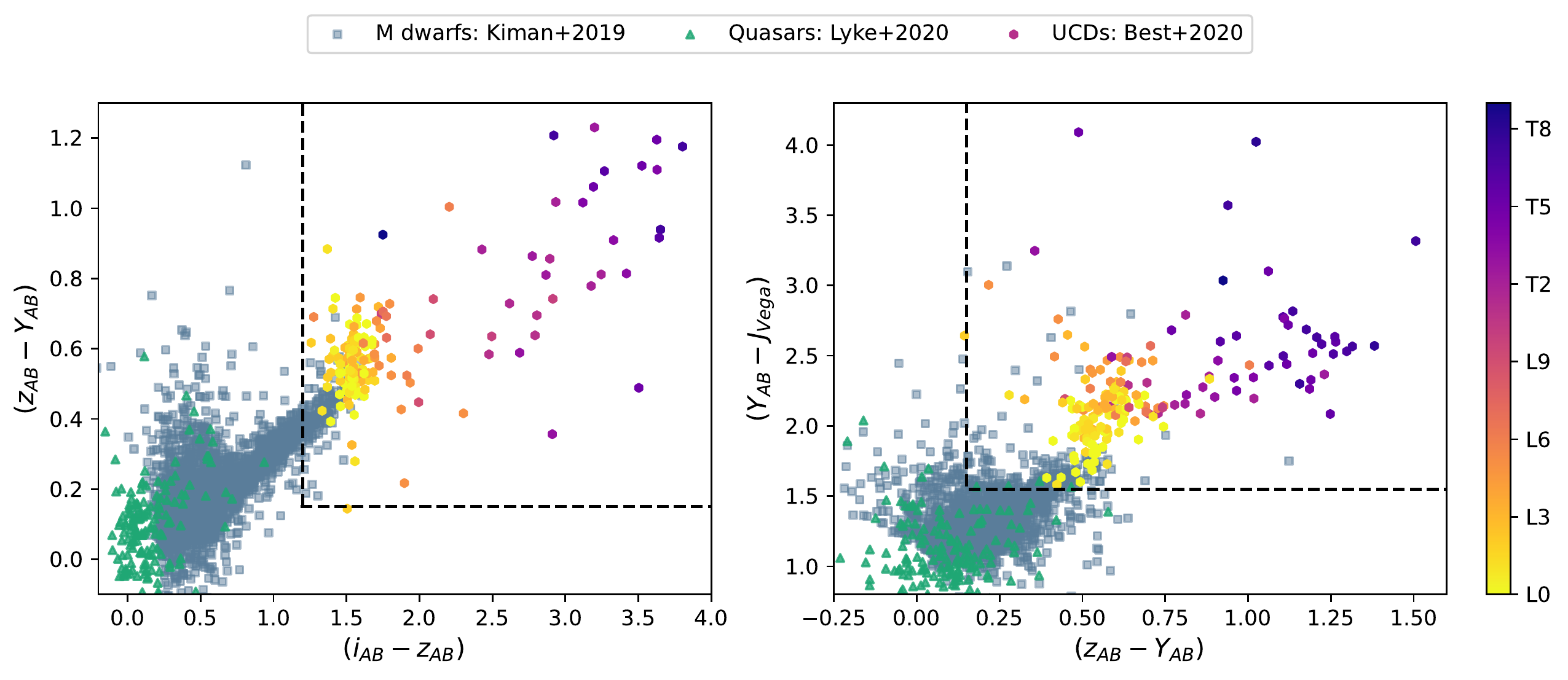}
\end{center}
\caption{Colour-colour diagrams for the M dwarfs \citep{Kiman2019} (blue squares), L and T dwarfs \citep{best2020}, shown as circles, and quasars with z > 4 \citep{Lyke2020} (green triangles). The colour coded represents objects with spectral type L0 and later. The black lines indicate the colour selection.}
\label{fig:calibration_samples}
\end{figure*}

The templates for the several colour indices as a function of the spectral type are shown in Figure \ref{fig:templates}. Also shown are the templates presented in \citet{carnero2019}. In comparison to our previous templates, there are no significant changes for the M and L dwarfs. For the T dwarfs, specially late types, color indices have changed typically by 0.1-0.2 mag, up to $\simeq 0.5$ mag in a couple of cases for T7 or later. This may be due to the clear increase in the number of objects that now contribute to the updated fit.
The redder $J-H$ and $H-Ks$ colours around L4 and T0 types are a known trend caused by the effect of condensate clouds and the variability in the clouds properties. Also, there is a blueward trend for T2 to T7 types in $J-H$, $H-Ks$ and $Ks-W1$ due to the loss of the cloud decks and the onset of CH$_{4}$ absorption. However, this trend diminishes for the latest types as very little flux remains to be absorbed by CH$_{4}$ \citep{Leggett2010}. The scatter for the later T types in $H-Ks$ and $Ks-W1$ is due to the variations in metallicity and gravity. The template colours are shown in Table \ref{tab:templates}. 

For the color selection of the UCDs, we follow the same methodology presented in \citet{carnero2019}. We analyze several color-color diagrams considering the UCDs and the contaminants samples presented earlier. The colour selection is meant to yield a sample of UCDs sources as complete as possible, at the expense of allowing some contamination by late-M dwarfs and extragalactic sources. The purity of our sample will be later improved using the photo-type classification (see section \ref{sec:classification}). We applied an optical band cut ($i-z$) $>$ 1.20, in order to remove the quasars, and also ($z-Y$) $>$ 0.15 and ($Y-J$) $>$ 1.55 to remove M dwarfs and other contamination sources. Figure \ref{fig:calibration_samples} shows the colour-colour diagrams where the colour selection was applied for known contaminants, M dwarfs and UCDs sources. Applying the color selection discussed above, the initial sample has 164,406 sources in DES+VHS+VIKING data. Among these, 76,184 objects have AllWISE W1 and W2 bands. The next step is to infer a photo-type for each object in the target sample. 

\section{Photo-type classification}
\label{sec:classification}

To infer a spectral type for objects in the target sample, we also closely follow the procedure described by \citet{carnero2019}, originally from \citet{skrzypek14}. The spectral type will be assigned by the minimization of the $\chi^2$ relative to our new empirical templates presented in Table \ref{tab:templates}. Only objects that have measurements in a minimum of $N_{bands}$= 6 bands (thus yielding 5 colour indices) are considered as having a reliable photo-type. We applied this minimum of six bands because we have observed a substantial improvement in photo-type determination with the number of filters available. The $\chi^2$ for the $k$-th source and the $j$-th spectral type is

$$\chi^2(\{m_{b}\},\{\sigma_{b}\},\hat{{m}}_{z,k,j},\{c_b\}) = \sum_{b=1}^{N_{bands}}{\left(\frac{m_{b,k}-\hat{{m}}_{z,k,j}-c_{b,j}}{\sigma_{b,k}}\right)}^2$$

\noindent where $m_{b,k}$ are the measured magnitudes for the source in all available filters, and $c_{b,j}$ are the template colors for the $j$-th spectral type and for the same bands. These latter are measured for all templates with respect to a reference band (in our case, the $z$ band). The $\sigma_{b,k}$ are the $k$-th source's photometric errors added in quadrature to the intrinsic scatter (from Section \ref{sec:templates_colour}). As for $\hat{{m}}_{z,k,j}$ in equation 2.1, it is the inverse variance weighted estimate of the reference magnitude, computed using all the source's magnitudes, their associated uncertainties and the given template colours for the $j$-th type, as follows

$$\hat{m}_{z,k,j}=\frac{\sum_{b=1}^{N_{bands}} {\frac{m_{b,k}-c_{b,j}}{\sigma_{b,k}^{2}}}} {\sum_{b=1}^{N_{bands}} {\frac{1}{\sigma_{b,k}^{2}}}}$$

\begin{figure}
\begin{center}
    \includegraphics[width=\columnwidth]{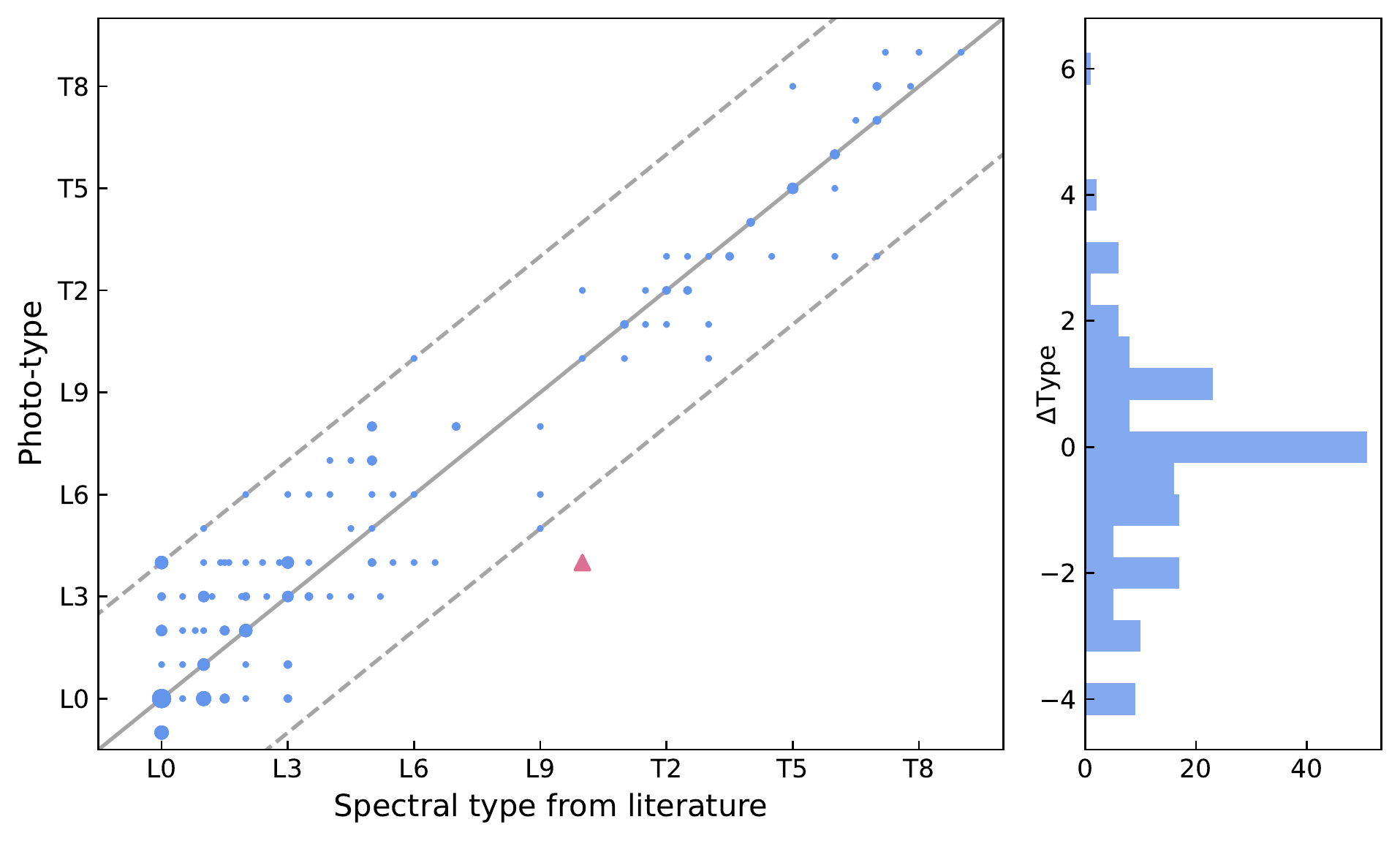}
\end{center}
\caption{Spectral classification from B2020 compilation against our photo-type classification. The dashed lines represent misclassification by four spectral types. The size of the circles scales as the cube of the number of repeated points. The histogram on the right shows the differences between the spectral types from the literature and our photo-types ($\Delta$Type).}
\label{fig:classif_best}
\end{figure}

\subsection{Comparison with the literature}

Figure \ref{fig:classif_best} shows the comparison between the spectral type from the literature and the photo-type method applied to the B2020 sample. As mentioned earlier, only objects with six or more valid magnitudes are shown. Only one object has a misclassification bigger than four spectral types: ULAS J223347+002214. However, this object is known as a strong binary candidate \citep{DayJones2013}.
The accuracy{\footnote{$\sigma = \frac{\sum_{j=1}^{N} {|\Delta Sp.T|}}{N} \frac{\sqrt{2\pi}}{2}$}} for the B2020 sample is $\sigma_{L}=1.7$ and $\sigma_{T}=1.1$ for L and T dwarfs, respectively. These values can be considered as an upper limit to the uncertainty in the assigned type. These values are compatible with those obtained by \citet{carnero2019} and \citet{skrzypek14}.
After testing the classification code, we obtain a photo-type for each object in our target sample. We used both DES+VHS+VIKING and DES+VHS+VIKING+AllWISE catalogs to estimate a photo-type. Our target sample now have 53,565 objects with photo-type $\geq$ L0 and six or more bands. 

Besides B2020, we also expect to recover in our target sample other UCDs candidates from the literature that are located in the DES footprint. As explained before, the colour selection was made considering objects that have spectroscopic confirmation, but these are currently limited in number. We thus benefit from assessing our sample selection by cross-matching our candidates to other sizeable samples of candidate sources, not only because of the increased numbers but also because this allows a direct comparison of different photo-types.

From the 1,361 objects presented by \citet{Skrzypek2016}, 154 are located in the DES footprint and 78 of them are present in our target sample. The missing 76 sources are due to three main reasons: i) a few objects are eliminated due to the colour selection and quality cuts applied to the DES data; ii) some are eliminated due to separation beyond 2$\arcsec$ match radius; iii) the main reason, however, is that most of them are eliminated because of our demand on availability of VHS+VIKING data.

\begin{figure}
\begin{center}
    \includegraphics[width=\columnwidth]{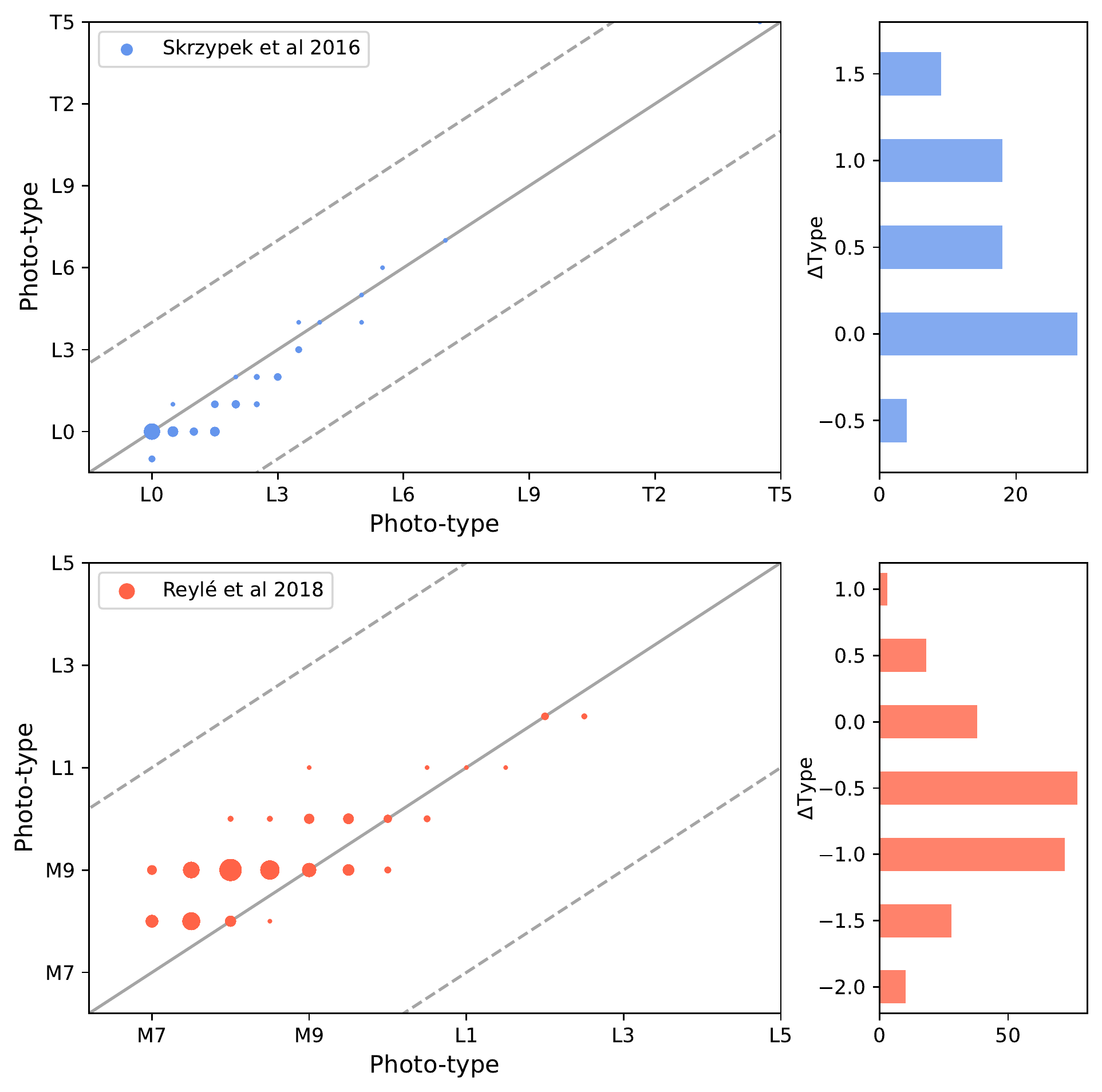}
\end{center}
\caption{Photo-type classification from \citet{Skrzypek2016} and \citet{Reyle2018} (x axis) and our photo-type classification (y axis). The dashed lines represent misclassification by four spectral types. The histogram on the right shows the differences between the photo-types from the literature and our photo-types ($\Delta$Type).}
\label{fig:classif_literature}
\end{figure}

\citet{Reyle2018} presented a sample of 14,915 $\geq$M7 and L candidates from the Gaia DR2 data, of which 2,224 are located in the DES footprint. However, only 40 of them are L dwarfs candidates and the remaining objects are M dwarfs. We end up with 248 of their objects in our target sample, 20 of which are L candidates and the remaining are M dwarfs (78 M7/M7.5, 102 M8/M8.5 and 48 M9/M9.5). The missing 20 L dwarfs were eliminated by either one of the reasons we mention above. The reduced number of M dwarfs in our sample is due to the color cuts imposed, as described in Section \ref{sec:templates_colour}. Figure \ref{fig:classif_literature} shows the comparison between the photo-types estimated from our classification code and those from these two other samples of UCD candidates. The median photo-type difference is of 0.5 for both \citet{Skrzypek2016} and \citet{Reyle2018} for objects with $z$ < 19. For fainter magnitudes we can only compare to \citet{Skrzypek2016} sample as \citet{Reyle2018} is limited in $z$ < 19 in our DES sample. For 19 < $z$ < 21 the median discrepancy is also 0.5.

\section{Towards the final sample: Target Validation}
\label{sec:sample_characteristics}


\subsection{Extragalactic contamination}

As in \citet{carnero2019}, we removed possible extragalactic contamination by running the \verb+Lephare+ photo-z code \citep{Arnouts1999, Ilbert2006} on the target sample using both a galaxy and quasar templates. We considered as extragalactic all sources that satisfied the following condition: $\chi_{Lephare}^{2} < \chi_{classif}^{2}$, where $\chi_{Lephare}^{2}$ and $\chi_{classif}^{2}$ are the best fit $\chi^2$ values from \verb+Lephare+ and from our photo-type code, respectively. 

From the 164,406 objects presented in our initial sample, only 53,565 have six or more bands and have a photo-type L0 and later. From this catalog of 53,565 L and T candidates, 34,116 were flagged as an extragalatic sources by \verb+Lephare+. Therefore, our final L and T dwarf candidate sample are constituted by a total of 19,449 objects. We also matched the 53,565 L and T dwarf candidates to SIMBAD \citep{Wenger2000} astronomical database in order to verify if the results provided by \verb+Lephare+ were in agreement with the literature. We found 327 objects in common, using a matching radius of 2$\arcsec$. From this list, only 63 were extragalactic sources and \verb+Lephare+ was able to discard 56. The 7 objects that remained in the sample were discarded. As discussed in \citet{carnero2019}, a residual contamination by extragalactic sources is estimated to be $\sim 5\%$.

We also tested running \verb+Lephare+ in the B2020 sample to verify the effect of the code on a pure UCD sample and only one object was flagged as an extragalactic: ULAS J222711-004547. ULAS J222711-004547 is known in the literature as a peculiar L dwarf. Since one ultracool dwarf was flagged as extragalatic by \verb+Lephare+ we decided to further investigate the 34,116 sources that were flagged as extragalatic sources using their proper motion information. In the next section we will discuss the details. 

\begin{figure}
\begin{center}
    \includegraphics[width=0.92\columnwidth]{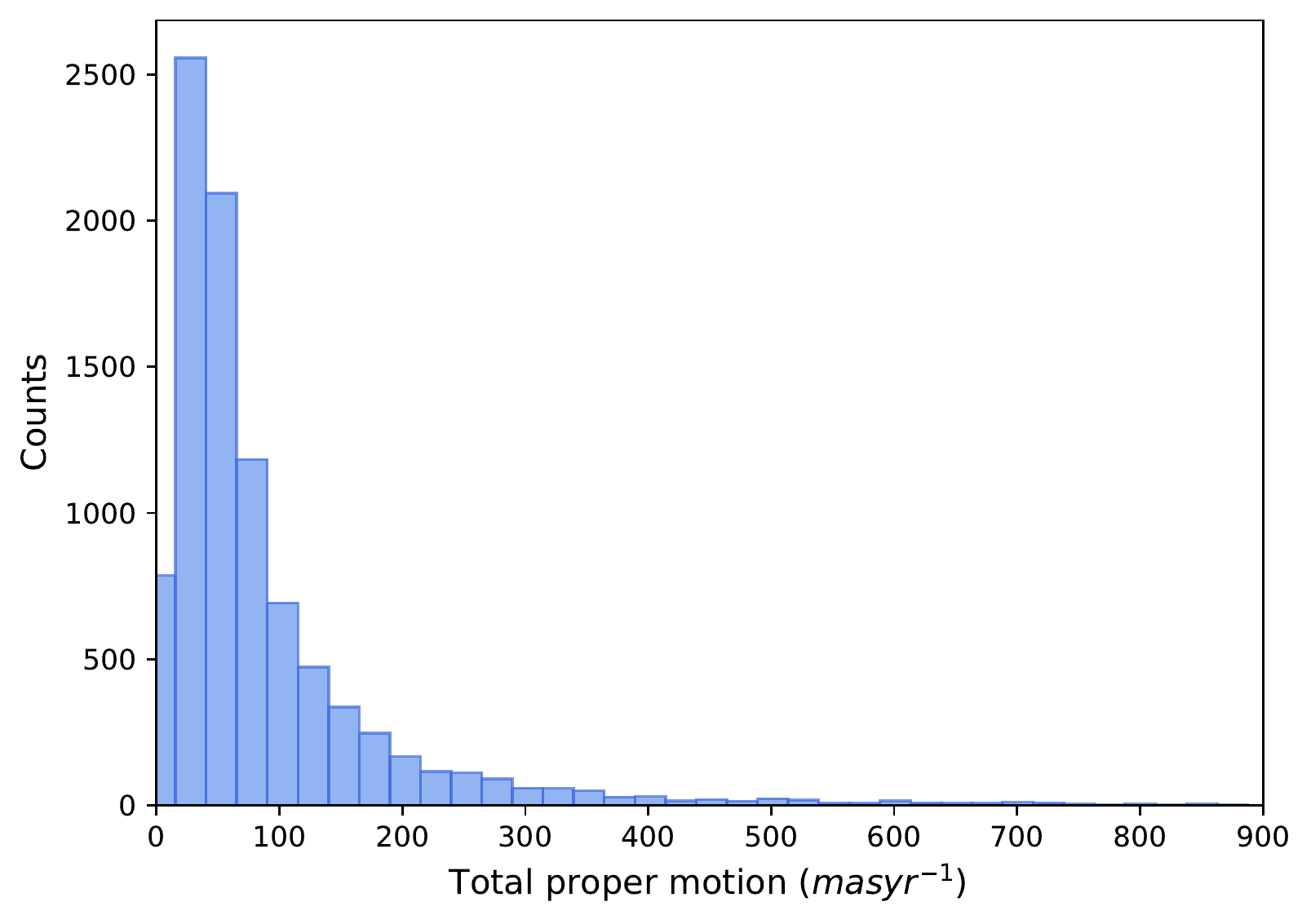}
\end{center}
\caption{Distribution of total proper motions for ultracool dwarf candidates in our sample. We only show here objects with well-measured proper motion according to our criteria presented in Section \ref{sec:pm}.}
\label{fig:ucds_pm_distribution}
\end{figure}

\begin{figure}
\begin{center}
    \includegraphics[width=\columnwidth]{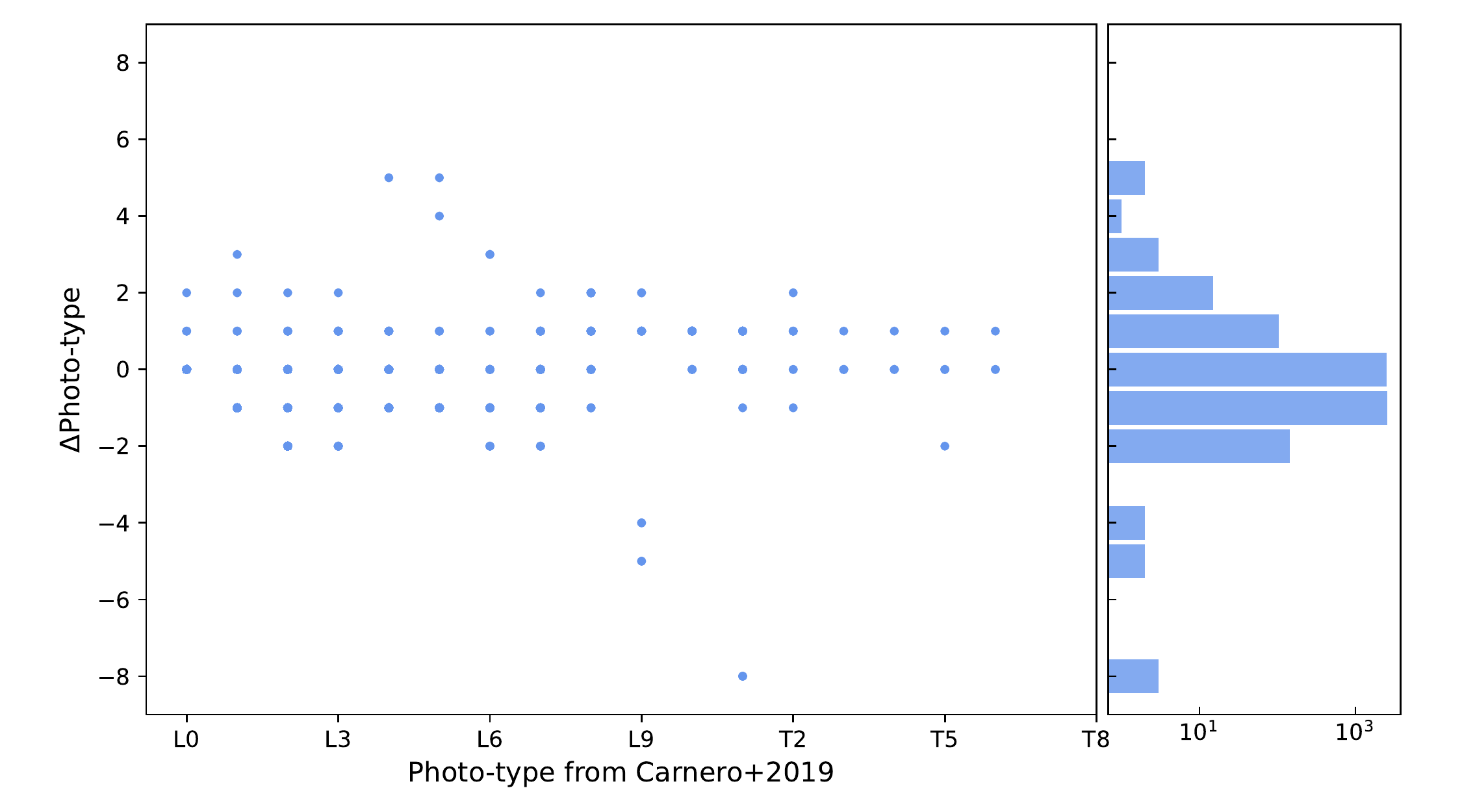}
\end{center}
\caption{Photo-type comparison between our new classification and the results from \citet{carnero2019} ($\Delta$Type). The histogram in the right shows that the vast majority of the objects have a difference of one spectral type, most now being classified as a slightly earlier type.}
\label{fig:photo_type_comparison}
\end{figure} 

\subsection{Proper Motion}
\label{sec:pm}

In addition to \verb+Lephare+, we used available proper motion catalogs in order to assess the Galactic or extragalactic nature of our candidate L and T dwarfs. If the source has a proper motion significantly different from zero, it is likely a Galactic one. We decided to use the proper motions from Gaia DR3 \citep{Gaia2022}, the CatWISE2020 catalog \citep{Marocco2021} and the NOIRLab Source Catalog (NSC) DR2 \citep{Nidever2021}. In particular, these last two catalogues extend towards faint enough magnitudes to cover a significant fraction (96\%) of our target sample of 53,565. These catalogs are responsible for several new discoveries, such as extreme T/Y subdwarfs \citep{Kirkpatrick2021, Meisner2021}, new ultracool dwarfs members of the Solar Neighbourhood \citep{Kota2022} and new wide binary systems \citep{Softich2022, Kiwy2022}. It is important to mention that in our sample of 53,565 L and T candidates, only 320 sources have Gaia DR3 proper motion measurements (this includes objects with RUWE $<$ 1.4 that ensures a good astrometric solution).  
 
We apply $\sigma_{\mu}/\mu$ $<$ 0.5 for all catalogs as a criterion for them to be considered Galactic sources. In the case of NSC, for some objects with large proper motion errors, $\sigma_{\mu} > $ 1,000 mas yr$^{-1}$, we felt the need to apply a more stringent selection criterion, $\sigma_{\mu}/\mu$ $<$ 0.1. Considering objects with Gaia DR3 proper motion measurements, for instance, only 12 out of the 320 sources are classified as an extragalactic source by \verb+Lephare+. However, 11 of them have proper motion from Gaia DR3 that satisfy our criteria. For the remaining objects flagged as extragalactic, 25,039 have proper motion measurements from CatWISE and NSC catalogs. In this case, 130 satisfies the criteria presented above. 
In total, 141 objects return to the L and T candidates sample. We conclude that proper motion data in conjunction with our adopted criteria do serve as a means to recover Galactic sources mistakenly classified as extragalactic by other means. Therefore, we have now 19,583 L and T dwarf candidates in the final sample. Figure \ref{fig:ucds_pm_distribution} shows the distribution of total proper motion ($\mu_{tot} = \sqrt{\mu_{\alpha cos\delta}^2+\mu_{\delta}^2}$) for the objects that satisfy the condition $\sigma_{\mu}/\mu$ $<$ 0.5 at least in one catalog (Gaia DR3, CatWISE2020 or NSC DR2). This sample has 9,278 objects.

\subsection{Comparison with our previous work}

From the objects presented in \citet{carnero2019}, 10,440 L and T dwarfs are present in the initial 164,406 sample of this paper (see Table \ref{tab:colorcuts}). The missing objects are due to a combination of slight changes in the DES footprint, the quality selection made in the target sample, changes in flags and photometric error criteria, and of lack of data in VHS+VIKING catalogs.

Imposing that the target must have six or more bands, something that was not applied in the past work, we end up with 8,512 in common. However, 5,342 objects are now classified as L or later. The remaining 3,170 are now classified as M9. This large migration across the M9/L0 border is expected due to the larger intrinsic scatter adopted here when compared to the previous work, as explained in Section \ref{sec:templates_colour}. Besides, we used the GalmodBD simulation code \citep{carnero2019} to estimate the reverse effect, namely  the contamination of M dwarfs to this new sample. We expect that $\sim 30\%$ of our sample is made up of late M dwarfs, the vast majority of them of M9 type. This is again somewhat larger than the 15-20\% estimated in our previous work. We should emphasize, however, that that this contamination is from sources of a very similar nature to our target L dwarfs. From the 5,342 L and T dwarfs still present on our sample, 24 were flagged as an extragalactic source either by \verb+Lephare+ or were listed in SIMBAD database. However, two flagged by \verb+Lephare+ have a proper motion measurement that satisfied our criteria. Therefore, in the end, 5,320 original L and T dwarfs from \citet{carnero2019} remain in the new sample presented here, while most of the missing ones are now classified as late M type. Figure \ref{fig:photo_type_comparison} shows the comparison between the photo-types from the previous work and those of the new candidate sample for objects in common. 

\begin{figure}
\begin{center}
    \includegraphics[width=0.92\columnwidth]{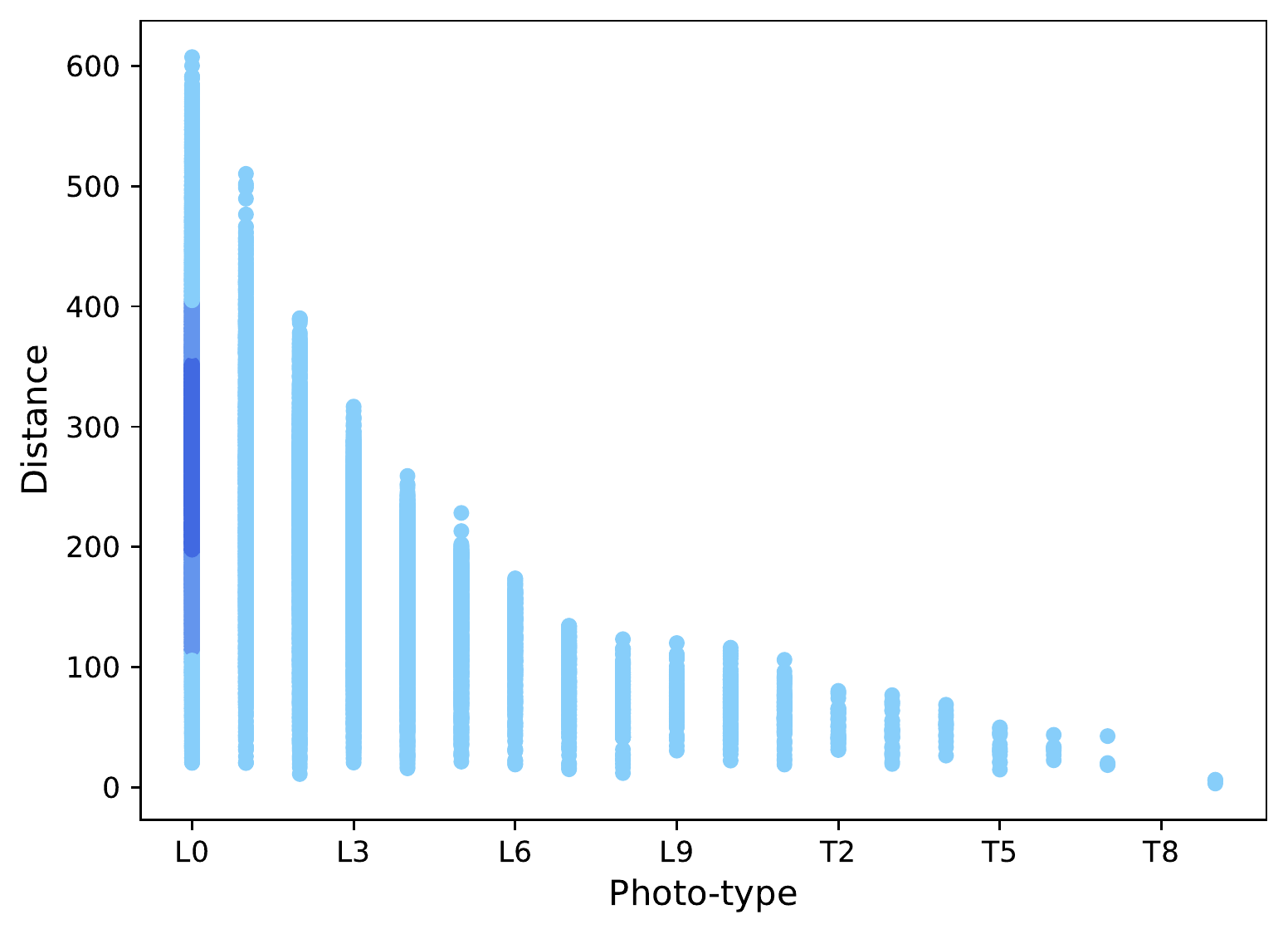}
\end{center}
\caption{Distances as a function of photo-type. Distances have been calculated using the average value from the distance modulus obtained using all available bands. The colour scale represents the density. Most ultracool dwarf candidates are early L at distances smaller than 650 parsecs.}
\label{fig:ucds_Y6_final}
\end{figure}

\subsection{New ultracool dwarf candidates}
\label{sec:new_candidates}

In total, 19,583 objects remain in our candidate sample, following all the criteria presented earlier. However, from this sample, 142 are included B2020, 5,257 additional ones were presented in \citet{carnero2019}, 26 from \citet{Reyle2018}, 5 from \citet{Skrzypek2016} and 54 from SIMBAD (mostly late-M dwarfs from other references, hence surveys, than those used here). There are 14,099 new UCD candidates. The table containing the ultracool dwarf candidates is available at: \url{https://des.ncsa.illinois.edu/releases/other/Y6-LTdwarfs}


Figure \ref{fig:ucds_Y6_final} shows the photo-type distribution vs photometric distance of the candidate sample of UCDs from this work. The final sample has only objects with six or more bands (used to estimate the photo-type), $\chi_{classif}^{2} < \chi_{Lephare}^{2}$ (or otherwise total proper motion significantly different from zero, if available) and a photo-type $\geq$L0. Here we see that this new sample is probing larger distances than those presented in \citet{carnero2019}. We now reach over 600 pc, while in our previous work we reached $\sim$ 480 pc.

We estimate photometric distances for our candidates following the same procedure explained in \citet{carnero2019}. We first calculate the absolute magnitudes for the UCD templates discussed in the previous chapter for all photometric bands and spectral types. We do that by using the template colours shown in Table \ref{tab:templates} and anchoring the absolute magnitude scale to the $M_{W2}$ values presented by \citet{dupuy12}. The distance for each UCD candidate in our sample is then determined from all its available apparent magnitudes and the template absolute magnitudes corresponding to its assigned spectral type. The mean distance over all available bands is assigned as the UCD distance. The distance uncertainty is obtained considering the photometric errors added in quadrature with the intrinsic scatter for each available band. We did not apply any correction for extinction, since this is expected to be small for the passbands we used and towards the relatively high Galactic latitudes covered by our sample. 

\begin{figure}
\begin{center}
    \includegraphics[width=0.95\columnwidth]{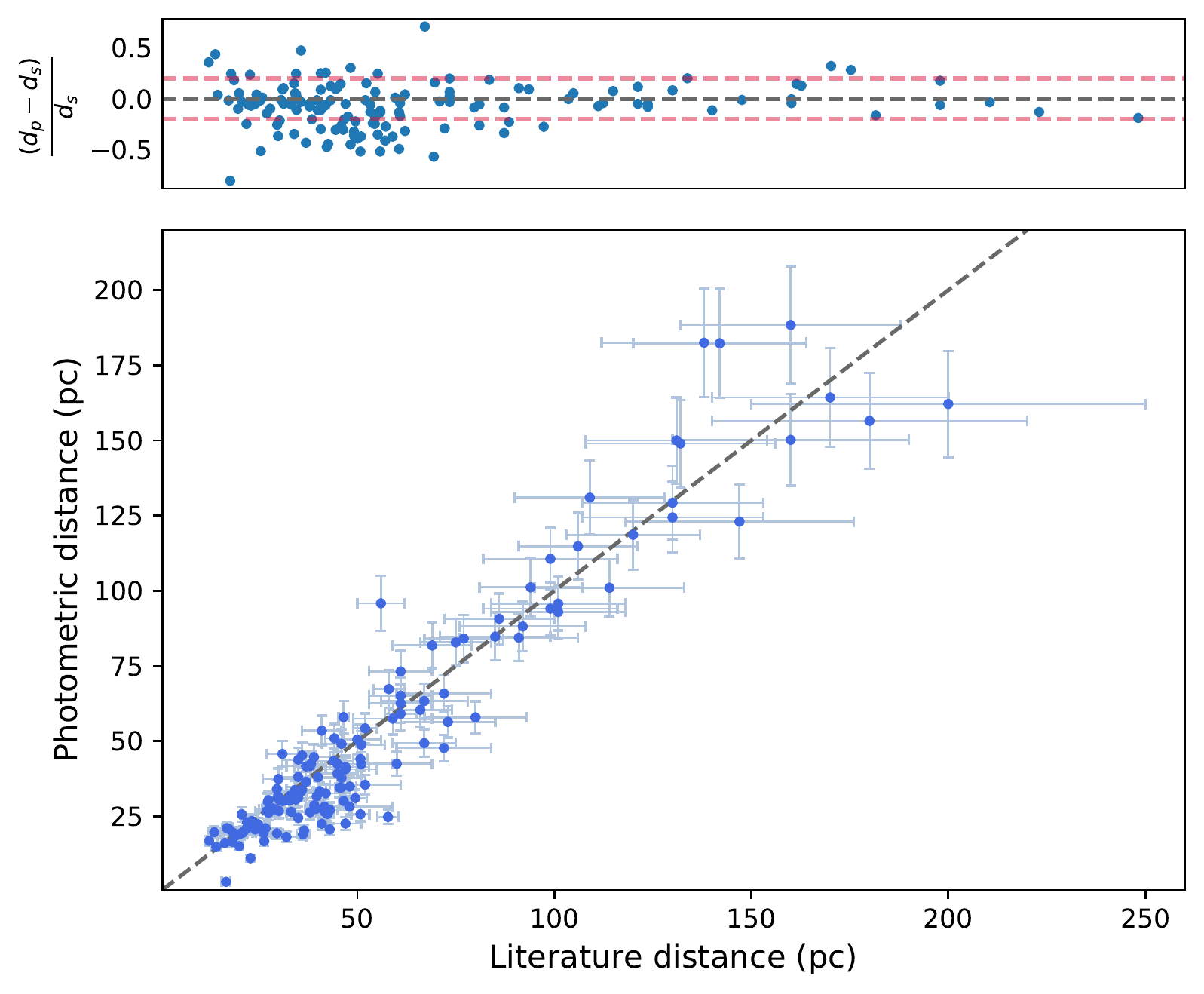}
\end{center}
\caption{Comparison between our photometric distances (d$_{p}$) and distances from the B2020 compilation (d$_{s}$), which has a mixture of trigonometric parallaxes and photometric distances. Our photometric distances tend to be slightly underestimated compared to those presented in B2020.}
\label{fig:comp_distance}
\end{figure}

We checked our photometric distances comparing with those presented in B2020, which comprehends several parallax and photometric measurements from the literature, as shown in Figure \ref{fig:comp_distance}. Our photometric distances tend to be slightly underestimated relative to those from B2020. This effect results from a tendency of assigning later types for the objects. Comparing our distance estimates and those from B2020 that have trigonometric parallax distances, the typical error in our photometric distances is $\sim$ 28\%. Also, the systematic offset seen in the Figure, in the sense of our distances being underestimated, is 18\% when we considered all objects from B2020, independent of the distance measurement method.

\section{UCD candidates catalogue applications}
\label{sec:sample_use}

\subsection{Young moving groups and association candidates}

Young moving groups and associations contain young stars ($\sim$ 10–100 Myr) and substellar objects whose similar kinematics imply that they originated in the same star-forming region. The members of a young association are a coeval population, where stars can serve as benchmarks to constrain metallicities and ages for substellar objects and to study models of star formation, for instance. Since our search targeted the general ultracool dwarfs population, we used the BANYAN $\Sigma$ code \citep{gagne2018} to estimate if any object in our sample is likely a moving group candidate member. The BANYAN $\Sigma$ algorithm uses a compiled list of bonafide members of 29 young moving groups and associations within 150 pc of the Sun and field stars within 300 pc to compute membership probability given the sky position, proper motion, distance, and radial velocity of targets using Bayesian inference. In our analyses, we divided the sample into: i) targets with Gaia DR3 information; ii) targets with CatWISE or NSC proper motion. For these latter, we also demanded that $\sigma_{\mu}/\mu$ $<$ 0.5. Also, we added in the samples radial velocity measurements from the literature when available.

\begin{table*}
\caption{Ultracool dwarfs new candidates and members recovered from the literature of young moving groups and associations. The objects in this table have proper motion and parallax from Gaia DR3 catalog.}
\centering
\begin{threeparttable}
\resizebox{\linewidth}{!}{%
\begin{tabular}{lccllllllcr}
\hline \hline
Object & RA & DEC & Ph.T & SpT & $\mu_{\alpha} cos \delta$  & $\mu_{\delta}$ & Parallax & RV & Prob. & Ref. \\
Name  & deg & deg &   &    & mas/yr & mas/yr & mas & km/s & & \\  
\hline
\multicolumn{11}{c}{\textbf{AB Doradus}} \\
\multicolumn{11}{l}{\tabitem Recovered Previously Candidate Members}  \\
J003256-440507 & 8.2335 & -44.0854 & L3 & L0 $\gamma$ & 127.84 $\pm$ 0.28 & -96.83 $\pm$ 0.31 & 28.95 $\pm$ 0.42 & - & 99\% & 2,3,5,6\\
J013847-345232 & 24.6981 & -34.8756 & L0 & - & 74.99 $\pm$ 0.47 & -52.01 $\pm$ 0.38 & 18.76 $\pm$ 0.63 & - & 93\% & 3 \\
J031645-284853 & 49.1886 & -28.8149 & L2 & L0 & 103.90 $\pm$ 0.22 & -94.71 $\pm$ 0.30 & 30.23 $\pm$ 0.34 & - & 98\% & 1,2,7 \\ 
J032529-431230 & 51.3728 & -43.2084 &  & M9 & 66.79 $\pm$ 0.26 & -20.77 $\pm$ 0.30 & 18.49 $\pm$ 0.25 & - & 88\% & 1 \\
J043350-421241 & 68.4578 & -42.2114 &  & M9* & 57.23 $\pm$ 0.26 & -29.37 $\pm$ 0.29 & 22.92 $\pm$ 0.24 & - & 93\% & 3 \\ 
J220645-421723 & 331.6883 & -42.2900 & L6 & & 128.67 $\pm$ 0.90 & -184.88 $\pm$ 0.93 & 34.08 $\pm$ 1.30 & - & 99\% & 1,2,7\\
\hline
\multicolumn{11}{c}{\textbf{$\beta$ Pictoris}} \\
\multicolumn{11}{l}{\tabitem Recovered Previously Candidate Members}  \\
J045327-175155 & 73.3605 & -17.8652 & L3 & L3 & 44.39 $\pm$ 0.38 & -20.60 $\pm$ 0.39 & 33.06 $\pm$ 0.54 & - & 99\% & 1,7,8 \\     
\hline
\multicolumn{11}{c}{\textbf{Carina}} \\ 
\multicolumn{11}{l}{\tabitem Recovered Previously Candidate Members} \\                     
J043531-644956 & 68.8773 & -64.8323 & L1 & M8.5 & 49.55 $\pm$ 1.33 & 36.32 $\pm$ 0.89 & 18.49 $\pm$ 0.61 & 19.7 $\pm$ 1.0  & 99\% & 1,7 \\  
\hline
\multicolumn{11}{c}{\textbf{Carina-Near}} \\ 
\multicolumn{11}{l}{\tabitem Recovered Previously Candidate Members} \\
J051929-450638 & 79.8699 & -45.1106 & L0 & L2* & 39.43 $\pm$ 0.91 & 66.55 $\pm$ 1.20 & 18.47 $\pm$ 0.98 & - & 96\% & 3 \\
\hline
\multicolumn{11}{c}{\textbf{Columba}} \\
\multicolumn{11}{l}{\tabitem New Candidate Members} \\
J051007-530626 & 77.5307 & -53.1072 & L0 & - & 26.57 $\pm$ 0.95 & 20.62  $\pm$ 1.31 & 11.18 $\pm$ 0.83 & - & 90\% & \\     
\multicolumn{11}{l}{\tabitem Recovered Previously Candidate Members} \\
J003443-410228 & 8.6798 & -41.0410 & L4 & L1 $\beta$ & 107.91 $\pm$ 0.80 & -59.28 $\pm$ 1.56 & 21.54 $\pm$ 1.18 & - & 82\% & 1,2,7 \\  
J050816-455751 & 77.0682 & -45.9641 & L0 & M8* & 25.33 $\pm$ 0.49 & 13.94 $\pm$ 0.64 & 11.64 $\pm$ 0.45 & - & 99\% & 3 \\
J051846-275646 & 79.6925 & -27.9460 & L4 & L1 $\gamma$ & 33.84 $\pm$ 0.51 & -4.82 $\pm$ 0.60 & 18.28 $\pm$ 0.59 & 24.35 $\pm$ 0.19 & 99\% & 2,5,6,7 \\ 
J055048-302006 & 87.6999 & -30.3351 & L1 & M9.4 & 20.46 $\pm$ 0.67 & -0.59  $\pm$ 0.85 & 18.50 $\pm$ 0.74 & 23.9 $\pm$ 1.4 & 99\% & 1,6 \\ 
J055538-413349 & 88.9064 & -41.5635 & L2 & L0.4 & 22.65 $\pm$ 0.83 & 15.76  $\pm$ 0.80 & 18.54 $\pm$ 0.65 & 23.5 $\pm$ 1.5 & 99\% & 1,6 \\ 
\hline
\multicolumn{11}{c}{\textbf{Tucana-Horologium Association}} \\ 
\multicolumn{11}{l}{\tabitem Recovered Previously Candidate Members} \\
J000658-643655 & 1.7423  & -64.6154 & L0 & & 86.20 $\pm$ 0.19 & -61.60 $\pm$ 0.21 & 23.17 $\pm$ 0.19 & - & 99\% & 1,6 \\
J003743-584624 & 9.4301  & -58.7732 & L4 & L0 $\gamma$  & 86.99 $\pm$ 0.91 & -49.95 $\pm$ 1.05 & 20.64 $\pm$ 0.81 & 6.62  $\pm$ 0.07 & 99\% & 2,4,6,7 \\     
J003815-640354 & 9.5629  & -64.0649 & L1 & M9.5 $\beta$ & 86.51 $\pm$ 0.30 & -47.71 $\pm$ 0.29 & 21.75 $\pm$ 0.27 & - & 99\%  & 2 \\ 
J011748-340327 & 19.4485 & -34.0574 & L3 & L1 $\beta$ & 108.27 $\pm$ 0.58 & -58.06 $\pm$ 0.71 & 25.40 $\pm$ 0.70 & - & 99\% & 2,4,6 \\
J012051-520036 & 20.2139 & -52.0099 & L4 & L1 $\gamma$ & 101.59 $\pm$ 0.89 & -44.85 $\pm$ 1.17 & 24.26 $\pm$ 0.94  & - & 99\% & 1,6,7 \\     
J014158-463358 & 25.4934 & -46.5661 & L4 & L0 $\gamma$ & 116.73 $\pm$ 0.35 & -46.62 $\pm$ 0.48  & 27.29 $\pm$ 0.44 & 6.41  $\pm$ 1.56 & 99\% & 6,7 \\  
J021039-301532 & 32.6612 & -30.2589 & L2 & L0 $\gamma$ & 101.63 $\pm$ 0.55 & -44.09 $\pm$ 0.51 & 24.65 $\pm$ 0.48 & 7.82  $\pm$ 0.27 & 99\% & 2,6,7 \\
J022155-541206 & 35.4799 & -54.2016 & L0 & M9 $\beta$  & 110.74 $\pm$ 0.20 & -21.91 $\pm$ 0.20 & 26.46 $\pm$ 0.19 & 10.18 $\pm$ 0.1  & 99\% & 2,4,6,7 \\     
J022355-581507 & 35.9786 & -58.2519 & L4 & L0 $\gamma$ & 105.22 $\pm$ 0.51 & -16.44 $\pm$ 0.50 & 25.17 $\pm$ 0.44 & 10.36 $\pm$ 0.23 & 99\% & 2,4,6,7 \\     
J022520-583730 & 36.3320 & -58.6250 & L0 & M9 $\beta$ & 100.88 $\pm$ 0.20 & -14.97 $\pm$ 0.20 & 24.25 $\pm$ 0.17 & 10.7 $\pm$ 2.2 & 99\% & 2,4,6,7 \\
J022657-532703 & 36.7365 & -53.4510 & L3 & L0 $\delta$ & 92.44 $\pm$ 0.47 & -18.78 $\pm$ 0.59 & 21.76 $\pm$ 0.43 & - & 99\% & 1,6,7 \\     
J023401-644207 & 38.5049 & -64.7020 & L2 & L0 $\gamma$ & 87.82 $\pm$ 0.65 & -4.93  $\pm$ 0.75 & 20.81 $\pm$ 0.62 & 11.76 $\pm$ 0.72 & 99\% & 2,4,5,6,7 \\
J024012-530553 & 40.0511 & -53.0980 & L0 & M9.5 & 96.30 $\pm$ 0.24 & -14.22 $\pm$ 0.28 & 23.49 $\pm$ 0.24 & 10.9 $\pm$ 2.2 & 99\% & 6,7 \\    
J024106-551147 & 40.2743 & -55.1963 & L4 & L1 $\gamma$ & 99.14 $\pm$ 0.86 & -13.3 $\pm$ 1.15 & 23.86 $\pm$ 0.80 & 11.73 $\pm$ 2.44 & 99\% & 1,7 \\ 
J024351-543220 & 40.9634 & -54.5388 & L0 & M9 & 91.71 $\pm$ 0.21 & -11.24 $\pm$ 0.22 & 22.29 $\pm$ 0.20 & 11.2 $\pm$ 2.2 & 99\% & 6,7 \\
J030149-590302 & 45.4545 & -59.0506 & L0 & M9 & 81.33 $\pm$ 0.17 & -2.01  $\pm$ 0.19 & 19.88 $\pm$ 0.15 & 12.3 $\pm$ 2.2 & 99\% & 6,7 \\    
J031143-323945 & 47.9273 & -32.6626 & L1 & M9.8* & 94.97 $\pm$ 0.43  & -24.90 $\pm$ 0.47 & 25.63 $\pm$ 0.56 & 10.6 $\pm$ 2.2 & 66\% & 1 \\
J032310-463124 & 50.7922 & -46.5232 & L4 & L0 $\gamma$ & 85.55 $\pm$ 0.87 & -7.32 $\pm$ 0.89 & 23.41 $\pm$ 0.70 & 13.0  $\pm$ 0.05 & 99\% & 2,5,6,7 \\
J035727-441731 & 59.3628 & -44.2918 & L2 & L0 $\beta$  & 76.69 $\pm$ 0.30 & -0.97 $\pm$ 0.41 & 21.28 $\pm$ 0.29 & 10.73 $\pm$ 4.6  & 99\% & 2,5,6,7 \\  
J044010-512654 & 70.0409 & -51.4484 & L3 & L0 $\gamma$ & 55.98 $\pm$ 1.23 & 18.87 $\pm$ 1.94 & 19.08 $\pm$ 1.15 & 15.6 $\pm$ 2.1 & 99\% & 1,6 \\ 
J045521-544616 & 73.8380 & -54.7710 & L0 & M9* & 54.09 $\pm$ 0.40 & 23.20 $\pm$ 0.45 & 19.09 $\pm$ 0.32 & - & 99\% & 3 \\     
J223536-590632 & 338.8989 & -59.1089 & L0 & M8.5 & 60.36 $\pm$ 0.20 & -84.16 $\pm$ 0.22 & 21.33 $\pm$ 0.23  & 2.9 $\pm$ 2.2 & 99\% & 2,6,7 \\     
J232253-615129 & 350.7216 & -61.8580 & L4 & L2 & 79.29 $\pm$ 0.80 & -80.17 $\pm$ 1.09 & 23.33 $\pm$ 0.96 & 6.75  $\pm$ 0.75 & 99\% & 1,5,6,7 \\ 
\hline\hline
\end{tabular}
}
\begin{tablenotes}
\small
\item \textbf{References:} (1) \citet{gagne2015}; (2) \citet{Faherty2016}; (3) \citet{gagne2018}; (4) \citet{Naud2017}; (5) \citet{Vos2019}; (6) \citet{Riedel2017}; \\ (7) \citet{Ujjwal2020}. 
\item \textbf{Notes:} *Photo-Type estimated using photometry.
\end{tablenotes}
\end{threeparttable}
\label{tab:moving_groups_gaia}
\end{table*}

\begin{table*}
\caption{Moving groups candidates with CatWISE2020 and NSC DR2 proper motion information. Candidates both new and recovered from the literature are listed.}
\centering
\begin{threeparttable}
\resizebox{\linewidth}{!}{%
\begin{tabular}{lllllllllllcr}
\hline \hline
Object & RA & DEC & Ph.T & SpT & $\mu_{\alpha} cos \delta$  & $\mu_{\delta}$ & $\mu_{\alpha} cos \delta$ & $\mu_{\delta}$ & Distance & RV & Prob. & Ref. \\
\hline
\multicolumn{13}{c}{\textbf{AB Doradus}} \\
\multicolumn{13}{l}{\tabitem New Candidate Members} \\  
J020047-510522 & 30.1975 & -51.0894 & L8 & - & 167.11 $\pm$ 4.40 & -85.81 $\pm$ 3.9 & 175.52 $\pm$ 2.28 & -68.00 $\pm$ 2.33 & 22.34 $\pm$ 2.06 & - & 99-99\% & \\  
J022609-161001 & 36.5383  & -16.1669 & L8 & - & 103.18 $\pm$  6.40 & -106.56 $\pm$  5.70 & 109.59 $\pm$  7.99 & -128.34 $\pm$  8.22 & 31.50 $\pm$ 2.92 & - & 98-99\% \\ 
J023618+004852 & 39.0753  &  0.8144 & L4 & - & 124.50 $\pm$  7.10 & -161.35 $\pm$  6.50 & 134.52 $\pm$  2.61 & -168.00 $\pm$  2.92 & 37.39 $\pm$ 3.45 & - & 92-75\% \\ 
J040232-264020 & 60.6320  & -26.6722 & T1 & - &  65.09 $\pm$ 24.00 &  -59.59 $\pm$ 25.60 & 80.81  $\pm$ 10.08 &  -66.82 $\pm$ 10.00 & 46.43 $\pm$ 4.40 & - & 90-81\% \\ 
J043250-562131 & 68.2098  & -56.3587 & L4 & - &  29.30 $\pm$  7.70 &   24.86 $\pm$  6.40 & 33.48  $\pm$  2.22 &   19.43 $\pm$  2.18 & 55.04 $\pm$ 5.09 & - & 85-95\% \\ 
J044842-592802 & 72.1762  & -59.4673 & L7 & - &  22.38 $\pm$  8.60 &   19.56 $\pm$  7.20 & 27.27  $\pm$  5.30 &   12.99 $\pm$  5.30 & 53.13 $\pm$ 4.94 & - & 94-95\% \\ 
J050656-251439 & 76.7333  & -25.2442 & L8 & - &  38.45 $\pm$  8.30 &  -61.19 $\pm$  8.30 & 36.50  $\pm$ 12.26 &  -64.88 $\pm$ 12.25 & 41.29 $\pm$ 3.84 & - & 98-98\% \\ 
J050928-311207 & 77.3671  & -31.2018 & L5 & - &  29.91 $\pm$ 10.00 &  -41.56 $\pm$  9.50 & 25.01  $\pm$  5.77 &  -32.87 $\pm$  5.87 & 50.23 $\pm$ 4.65 & - & 96-97\% \\ 
J051244-502007 & 78.1825  & -50.3351 & L4 & - &  43.30 $\pm$  5.40 &   14.75 $\pm$  4.90 & 43.73  $\pm$  2.47 &   19.18 $\pm$  2.39 & 35.33 $\pm$ 3.26 & - & 99-98\% \\ 
J052114-373332 & 80.3095  & -37.5590 & L5 & - &  10.21 $\pm$  7.20 &  -35.93 $\pm$  7.19 & 15.46  $\pm$  2.46 &  -40.04 $\pm$  2.52 & 35.79 $\pm$ 3.31 & - & 98-99\% \\ 
J053808-493406 & 84.5327  & -49.5683 & L2 & - &  26.25 $\pm$  7.60 &    1.05 $\pm$  7.30 & 13.32  $\pm$  1.86 &   -2.46 $\pm$  1.88 & 58.22 $\pm$ 5.38 & - & 81-94\% \\
\multicolumn{13}{l}{\tabitem Recovered Previously Candidate Members}  \\
J032642-210208 & 51.6765 & -21.0356 & L8 & L5 $\beta$ &  83.09 $\pm$ 5.00 & -143.70 $\pm$ 4.50 &  90.57 $\pm$ 3.42 & -144.60 $\pm$ 3.13 & 19.48 $\pm$ 1.80 & 22.91 $\pm$ 20.07 & 65-89\% & 1,2,5,6 \\
J040627-381210 & 61.6117 & -38.2028 & L4 & L0 $\gamma$ & 37.72 $\pm$ 7.90 & -11.14 $\pm$ 7.50  & 41.90 $\pm$ 5.38  & 1.90 $\pm$ 5.53 & 47.70 $\pm$ 4.41 & - & 66\% & 2,5 \\
J041352-401009 & 63.4646 & -40.1692 & L4 & L2.5 & 53.82 $\pm$ 7.9 & -10.1 $\pm$ 7.1 & 39.45 $\pm$ 3.17 & 4.22 $\pm$ 3.19 & 47.56 $\pm$ 4.39 & - & 76\% & 1 \\
\hline
\multicolumn{13}{c}{\textbf{$\beta$ Pictoris}} \\
\multicolumn{13}{l}{\tabitem New Candidate Members} \\  
J045544-250107 &  73.9353 & -25.0187 & L5 & - &   43.48 $\pm$  8.20 &  -0.22 $\pm$  8.00 &  31.09 $\pm$  4.78 &  -9.96 $\pm$  4.95 & 41.72 $\pm$ 3.86 & - & 89-98\% \\  
J202436-544944 & 306.1502 & -54.8289 & L8 & - &   37.74 $\pm$ 10.60 & -86.00 $\pm$  9.90 &  53.44 $\pm$ 11.51 & -82.63 $\pm$ 11.65 & 42.28 $\pm$ 3.94 & - & 88-98\% \\
J213422-582853 & 323.5926 & -58.4814 & L8 & - &   61.87 $\pm$ 14.10 & -93.82 $\pm$ 13.00 &  85.69 $\pm$  5.11 & -88.91 $\pm$  5.22 & 43.14 $\pm$ 4.01 & - & 62-92\% \\ 
\multicolumn{13}{l}{\tabitem Recovered Previously Candidate Members}  \\
J034209-290432 & 55.5391 & -29.0755 & L3 & L0 $\beta$ & 37.26 $\pm$ 7.2 & 0.44 $\pm$ 6.6 & 60.95 $\pm$ 2.78 & -7.79 $\pm$ 2.90 & 42.31 $\pm$ 3.91 & - & 93\% & 1,2,5 \\ 
J053620-192040 & 84.0834 & -19.3445 & L6 & L2 $\gamma$ & 31.73 $\pm$ 5.50 & -13.04 $\pm$ 5.60 & 33.24 $\pm$ 2.48 & -18.65 $\pm$ 2.54 & 22.58 $\pm$ 2.08 & 22.06 $\pm$ 0.70 & 99-98\% & 2,4,5 \\ 
\hline
\multicolumn{13}{c}{\textbf{Carina-Near}} \\
\multicolumn{13}{l}{\tabitem New Candidate Members} \\ 
J033555-443916 & 53.9784  & -44.6545 & T5 & - & 161.37 $\pm$ 17.90 & 133.76 $\pm$ 17.60 & 209.40 $\pm$ 6.07 & 125.57 $\pm$ 5.88 & 27.03 $\pm$ 2.59 & - & 86-95\% \\ 
J042013-253924 & 65.0526  & -25.6567 & L6 & - &  90.09 $\pm$  9.30 &  71.09 $\pm$  9.20 &  95.23 $\pm$ 4.99 &  63.46 $\pm$ 4.85 & 42.72 $\pm$ 4.82 & - & 79-92\% \\
\hline
\multicolumn{13}{c}{\textbf{Columba}} \\
\multicolumn{13}{l}{\tabitem New Candidate Members} \\  
J043838-460256 & 69.6573 & -46.0488 & L5 & - &  38.75 $\pm$  8.20 &  7.94 $\pm$  7.60 &  39.61 $\pm$  2.85 &  14.01 $\pm$  2.88 & 48.05 $\pm$ 4.44 & - & 67-94\% \\ 
\multicolumn{13}{l}{\tabitem Recovered Previously Candidate Members}  \\
J031819-643322 & 49.5772  & -64.5561 & L2 & - &  48.75 $\pm$  7.80 & 9.05 $\pm$ 7.00 & 48.62  $\pm$  2.32 & 8.18 $\pm$ 2.28 & 59.03 $\pm$ 5.46 & 12.6 $\pm$ 2.4 & 54\% & 1\\ 
J041859-450741 & 64.7453 & -45.1281 & L4 & L3 $\gamma$ & 60.44 $\pm$ 5.90 & 15.26 $\pm$ 5.00 & 58.56 $\pm$ 2.65 & 8.00 $\pm$ 2.51 & 38.00 $\pm$ 3.51 & 15.1 $\pm$ 2.1 & 83-92\% & 2,4,5 \\
J051050-184356 & 77.7070 & -18.7321 & L3 & L2 $\beta$ & 81.77 $\pm$ 5.80 & -50.12 $\pm$ 6.1 & 83 $\pm$ 3.01 & -44.10 $\pm$ 2.98 & 31.86 $\pm$ 2.94 & 23.2 $\pm$ 1.3 & 97-55\% & 1,2 \\ 
J054008-364218 & 85.0345 & -36.7050 & L4 & L2.3 & 21.47 $\pm$ 6.9 & -3.23 $\pm$ 6.8 & 29.45 $\pm$ 2.52 & 4.49 $\pm$ 2.56 & 38.17 $\pm$ 3.53 & - & 99\% & 1 \\ 
\hline
\multicolumn{13}{c}{\textbf{Octans}} \\
\multicolumn{13}{l}{\tabitem New Candidate Members} \\
J005503-533413 & 13.7636  & -53.5703 & L0 & - &  31.60 $\pm$  9.30 & 28.61 $\pm$  8.40 & 29.23 $\pm$ 1.58 & 16.56 $\pm$ 1.67 & 77.06 $\pm$ 7.13 & - &  96-71\% \\         
\hline
\multicolumn{13}{c}{\textbf{Tucana-Horologium Association}} \\
\multicolumn{13}{l}{\tabitem New Candidate Members} \\
J024725-492032 & 41.8555  & -49.3423 & L8 & - & 109.25 $\pm$  7.60 &  -16.97 $\pm$  6.60 & 106.75 $\pm$ 10.08 & -28.82 $\pm$ 10.09 & 50.02 $\pm$ 4.71 & - & 92-80\% \\ 
\multicolumn{13}{l}{\tabitem Recovered Previously Candidate Members}  \\
J203345-563535 & 308.4365 & -56.5931 & L3 & L0 $\gamma$ & -3.95 $\pm$ 6.7 & -74.19 $\pm$ 6.6 & 13.69 $\pm$ 2.47 & -84.69 $\pm$ 2.72 & 40.66 $\pm$ 43.77 & - & 56\% & 1,2,5 \\ 

\hline\hline
\end{tabular}}
\begin{tablenotes}
\small
\item \textbf{References:} (1) \citet{gagne2015}; (2) \citet{Faherty2016}; (3) \citet{Naud2017}; (4) \citet{Vos2019}; (5) \citet{Riedel2017}; (6) \citet{Ujjwal2020}
\end{tablenotes}
\end{threeparttable}
\label{tab:moving_groups_cat_nsc}
\end{table*}

\begin{table*}
\centering
\caption{The common proper motion and distance pair candidates identified among the UCD sample. The ID in $Jhhmm \pm ddmm$ format is based on the primary coordinates and the letters A and B represent a different UCD. The flag indicates 0=common distance and common PM, the latest based on all available catalogs, 0=common distance and common proper motion according to CatWISE2020 and 0=common distance and common proper motion according to NSC DR2.}
\label{tab:comoving_systems}
\begin{threeparttable}
\resizebox{\linewidth}{!}{
\begin{tabular}{lllllcclllllllcr}
\hline\hline
\multicolumn{1}{c}{ID} & \multicolumn{4}{c}{Position (deg)} & \multicolumn{2}{c}{Ph.T} &  \multicolumn{4}{c}{Proper Motion (mas/yr)} & \multicolumn{2}{c}{Distance (pc)} & Sep ($\arcsec$) & Ref. & Flag \\
& $\alpha_{A}$ & $\delta_{A}$ & $\alpha_{B}$ & $\delta_{B}$ & A & B & $\mu_{\alpha} cos \delta$  & $\mu_{\delta}$ & $\mu_{\alpha} cos \delta$  & $\mu_{\delta}$ & $d_{A}$ & $d_{B}$ &  & \\
\hline
\multicolumn{16}{l}{\tabitem New Candidate Systems} \\
J004316-320343 & 10.815 & -32.062 & 10.800 & -32.110 & T2 & L9 & -11.17 $\pm$ 18.00 & 46.72 $\pm$ 17.00 & -19.65 $\pm$ 14.20 & 51.60 $\pm$ 13.80 & 41.86 $\pm$ 3.93 & 41.34 $\pm$ 3.85 & 176.5 & 5 & 101 \\ 
J020903-124420 & 32.262 & -12.739 & 32.262 & -12.738 & L0 & L0 & 2.45 $\pm$ 4.78 & 22.89 $\pm$ 5.15 & 0.69 $\pm$ 4.92 & 33.54 $\pm$ 5.33 & 155.46 $\pm$ 14.81 & 151.94 $\pm$ 14.45 & 54.2 & 5 & 110 \\ 
J022636-013744 & 36.651 & -1.629 & 36.651 & -1.629 & L0 & L0 & 6.05 $\pm$ 4.04 & -38.40 $\pm$ 4.04 & 0.79 $\pm$ 4.01 & -37.45 $\pm$ 4.01 & 166.28 $\pm$ 16.12 & 168.59 $\pm$ 16.37 & 2.4 & 5 & 110 \\ 
J030422-135839 & 46.090 & -13.977 & 46.049 & -14.016 & L6 & L8 & 109.39 $\pm$ 13.8 & 45.14 $\pm$ 13.4 & 132.71 $\pm$ 36.1 & 32.35 $\pm$ 37.9 & 55.30 $\pm$ 5.15 & 90.09 $\pm$ 9.53 & 199.4 & 5 & 000 \\ 
\multicolumn{15}{l}{\tabitem Recovered systems} \\
J013036-444542 & 22.648 & -44.761 & 22.649 & -44.761 & L8 & L0 & 124.98 $\pm$ 3.84 & -33.63 $\pm$ 4.83 & 116.31 $\pm$ 1.44 & -27.87 $\pm$ 1.74 & 27.00 $\pm$ 2.50 & 27.84 $\pm$ 2.57 & 3.1 & 2 & 110 \\ 
J014611-050851 & 26.547 & -5.147 & 26.546 & -5.147 & L4 & L7 & 81.78 $\pm$ 1.98 & -218.51 $\pm$ 1.93 & 80.51 $\pm$ 4.84 & -214.01 $\pm$ 4.75 & 29.51 $\pm$ 2.73 & 45.07 $\pm$ 4.18 & 3.2 & 3 & 110 \\ 
J055146-443411* & 87.941 & -44.569 & 87.941 & -44.570 & L0 & L0 & -61.01 $\pm$ 1.38 & -16.71 $\pm$ 1.44 & -61.02 $\pm$ 0.88 & -13.07 $\pm$ 0.83 & 76.64 $\pm$ 7.08 & 66.20 $\pm$ 6.12 & 2.2 & 1 & 111 \\ 
J231349-455025 & 348.455 & -45.840 & 348.455 & -45.841 & L5 & L4 & 53.76 $\pm$ 12.33 & 6.06 $\pm$ 12.36 & 55.02 $\pm$ 4.94 & 13.36 $\pm$ 5.0 & 106.25 $\pm$ 10.25 & 81.25 $\pm$ 7.62 & 4.1 & 4 & 000 \\ 
\hline \hline
\end{tabular}
}
\begin{tablenotes}
\small
\item \textbf{References:} (1) \citet{Billers2005}; (2) \citet{Dhital2011} (3) \citet{Softich2022}; (4) \citet{dalponte2020}; (5) This work 
\item \textbf{Notes:} *Proper motion from Gaia DR3.
\end{tablenotes}
\end{threeparttable}
\end{table*}

It is important to mention that we ran BANYAN $\Sigma$ twice if the object has CatWISE2020 and NSC DR2 proper motion. In this case, we only kept candidates whose BANYAN results were the same. We found that 60 objects among our list were already reported in the literature as moving groups candidate members. The recovered members are shown in Table \ref{tab:moving_groups_gaia} and Table \ref{tab:moving_groups_cat_nsc} along with new candidates. Table \ref{tab:moving_groups_gaia} contains only the objects with proper motion and parallax from Gaia DR3. Table \ref{tab:moving_groups_cat_nsc} contains objects with proper motion from CatWISE2020 and NSC DR2, photometric distances, and BANYAN $\Sigma$ probabilities according to the catalog used. We also added in both tables the spectral type from the literature (when available) besides our photo-type. The objects from the literature have added to their spectral type the gravity subtypes $\alpha$, $\beta$, and $\gamma$ to designate objects of normal gravity, intermediate gravity, and very low gravity, respectively. Also, the $\delta$ suffix denotes objects with an even younger age (typically less than a few Myr) and lower surface gravity than those associated with the $\gamma$ suffix \citep{Kirkpatrick2006}.

The young moving groups candidates (new and recovered) that we found are: 20 in AB Doradus \citep[AB Dor, 110-150 Myr;][]{Luhman2005, Barenfeld2013}, six in $\beta$ Pictoris \citep[$\beta$ Pic, 22 $\pm$ 6 Myr;][]{Shkolnik2017}, 11 in Columba \citep[Col, 42 $^{+6}_{-4}$ Myr;][]{Bell2015}, one in Carina \citep[Car, 45 $^{+11}_{-7}$ Myr;][]{Bell2015}, three in Carina-Near \citep[CarN, 200 $\pm$ 50 Myr;][]{Zuckerman2006}, one in Octans \citep[OCT, 35 $\pm$ 5 Myr;][]{Murphy2015} and 25 in Tucana-Horologium \citep[THA, 45 $\pm$ 4 Myr;][]{Bell2015}. We did not include any candidate member from Argus association considering its high level of contamination \citep{Bell2015}. 

We found 20 new candidate members to young associations with Bayesian membership probability above 90\%, at least in one catalog. For objects in common with the literature, we analysed each case individually considering not only the difference in kinematics between this work and previous ones (our work probably making use of more recent and robust proper motion measurements), but also the use of BANYAN $\Sigma$ (more recent and updated code) results in substitution to those presented by BANYAN II or BANYAN I, for instance. 12 objects are now classified as field members according to our results and are not presented in the following tables. The ambiguous objects were placed in the group indicated by our BANYAN $\Sigma$ run. The objects with discrepancies in the classification are:

\begin{itemize}
    \item \textit{J003443-410228}: this object was first presented in \citet{Faherty2016} and more recently in \citet{Ujjwal2020} as a THA candidate member. However, using proper motion and parallax measurements from Gaia DR3, we identified it as Col candidate member (83\% of probability).
    \item \textit{J031645-284853}: \citet{Faherty2016} presented as ambigous AB Dor member by BANYAN II. More recently, \citet{Ujjwal2020} classified as THA candidate member. Here using Gaia DR3 proper motion and parallax we classified as AB Dor candidate member (98\% of probability).
    \item \textit{J040627-381210}: \citet{Riedel2017} classified this object as field member using the LACEwING code. However, \citet{Faherty2016} presented as Col candidate member using BANYAN II, Octans by LACEwING and field object by BANYAN I. Here, the use of NSC proper motion into BANYAN $\Sigma$ also indicated as field object (66\% probability).
    \item \textit{J041352-401009}: \citet{gagne2015} classified this object as Col candidate member. The BANYAN $\Sigma$ classified as $\beta$ Pic member when we used the NSC DR2 proper motion (32\% probability) and as AB Dor member using CatWISE2020 data (76\% of probability).
    \item \textit{J034209-290432}: \citet{gagne2015} and \citet{Riedel2017} classified this object as THA candidate member. Here, BANYAN $\Sigma$ classified as field member when we used the CatWISE proper motion (94\% probability) and as a $\beta$ Pic using NSC DR2 (93\% probability).
    \item \textit{J041859-450741}: \citet{Faherty2016} presented as ambiguous THA member by BANYAN II and AB Dor candidate by LACEwING and BANYAN I. Both \citet{Vos2019} and \citet{Riedel2017} identified this object as AB Dor candidate member. Here, BANYAN $\Sigma$ identified as Col candidate member.
    \item \textit{J031819-643322}: \citet{gagne2015} presented this object as THA member. Here, BANYAN $\Sigma$ classified as field using NSC proper motion (99\% probability) and Col member using CatWISE2020 (53\% probability).
    \item \textit{J054008-364218}: \citet{gagne2015} classified this object as Col candidate member. Here, BANYAN $\Sigma$ classified as $\beta$ Pic member when we used the CatWISE2020 proper motion (76\% probability) and as Col member using NSC DR2 data (99\% probability). 
    \item \textit{J203345-563535}: \citet{gagne2015}, \citet{Faherty2016} and \citet{Riedel2017} presented this object as THA member. Here BANYAN $\Sigma$ classified as field member when we used the CatWISE2020 proper motion (99\% probability).
\end{itemize}

Despite the recovered and new candidate members to younger populations, still the vast majority of 99.1\% of our sample that has significant proper motion is composed of field objects. Also, it is important to mention that the comparison between our photo-type estimate and spectral type from candidate members of young associations from the literature shows a systematic discrepancy of up to +4 types in some cases. This may be the effect of deviant colors attributed to enhanced dust or thick photospheric clouds, that shift the flux to longer wavelengths in young objects \citep{Faherty2016}.

\subsection{Variability}

Photometric variability can help to understand atmospheric inhomogeneities in ultracool dwarfs, as it is sensitive to the spatial distribution of condensates as the object rotates. It has been studied in the more massive field L and T dwarfs, but still the variability of the younger and low-gravity objects is less understood. For instance, only a small sample of variability in low-gravity objects \citep{Metchev2015, Vos2019} has been detected so far.

\begin{figure*}
\begin{center}
    \includegraphics[width=0.98\linewidth]{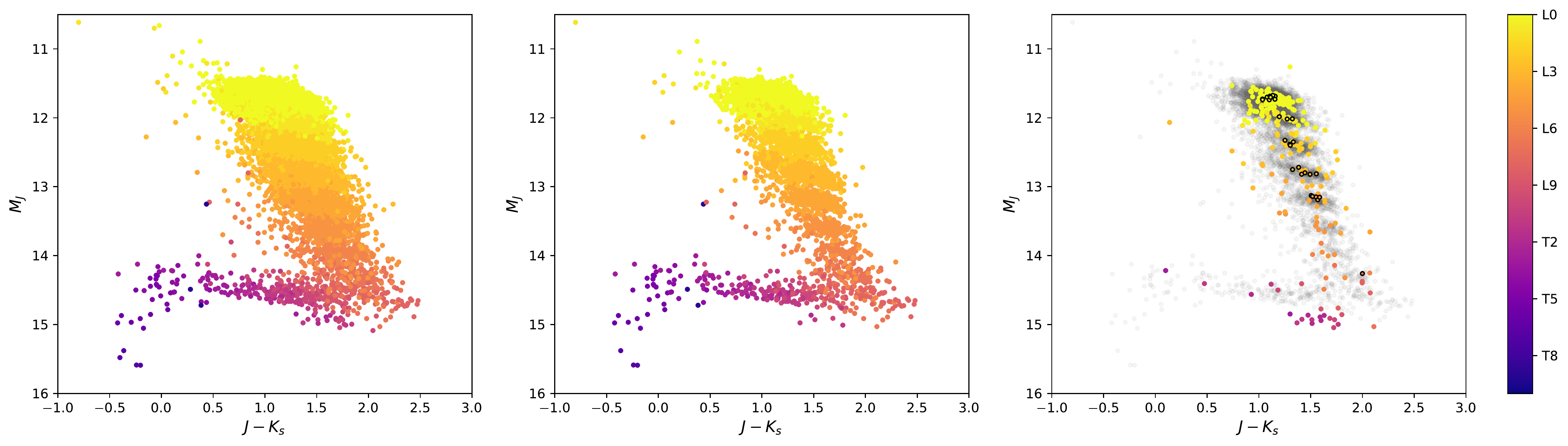}
\end{center}
\caption{Colour $J-K_{s}$ vs absolute magnitude in J band. The left panel shows all the 19,583 ultracool dwarf candidates. The mid panel shows only ultracool dwarfs candidates with significant proper motion. The right panel shows the photometric variable objects identified in the Dark Energy Survey Y6 variability catalog. The points with black contour represent the 28 young candidate objects among the variable sample. The shaded background is made up of all sources from the middle panel.}
\label{fig:variability}
\end{figure*}

Here, we first used DES Y6 variability catalog described in \citet{Stringer2021} to search for variable sources among our 19,583 ultracool candidates sample and we found 291 of those. There are several available statistics to select variable sources in this particular catalog. The reduced $\chi^{2}$ (\verb+RED_CHISQ_PSF_grizy+ $\geq$ 3.3) seems the most efficient to separate variable objects, for instance RR Lyrae, from standard stars. From these 291 variable candidate sources, 130 are also in the Transiting Exoplanet Survey Satellite (TESS) Input Catalog \citep{Stassun2018}, 28 are young objects already identified in the literature and that were discussed in the previous section and are presented in Table \ref{tab:moving_groups_gaia} and Table \ref{tab:moving_groups_cat_nsc}.
It is out of the purpose of this work to further analyze in detail these variable candidates. However, this type of sample may be a starting point for studies regarding the cloud formation and dissipation on brown dwarf atmospheres and to assess if low-gravity objects are more likely variable than their field ultracool dwarfs counterparts \citep{Metchev2015}.

Figure \ref{fig:variability} shows colour-magnitude diagrams for the entire sample of ultracool dwarf candidates presented in Section \ref{sec:sample_characteristics} (left panel), for ultracool dwarf candidates that have significant proper motion (satisfy the criteria from Section \ref{sec:pm}; middle panel) and for the photometrically variable candidates identified in the DES variability catalog (right panel). From these latter, the 28 young candidate objects mentioned above are highlighted. Absolute magnitudes were calculated using our photometric distance estimates. The variable sources seem to roughly follow the same colour-magnitude properties as our full sample of ultracool dwarfs. We may not see subtle redder colours for the highlighted young L types because according to our methodology we tend to attribute later spectral types for young objects, as mentioned in the previous section. Here, 10\% of the young (L0-L7) candidate members to moving groups show photometric variability, a lower fraction when compared to \citet{Vos2019} that found 30$^{+16}_{-8}$\% for the frequency of variable young objects in L0–L8.5 spectral type range. The remaining variable objects span the L0-T3 range of photo-types. We find that they correspond to 1.3\% of the total populations in the range L0-L8, and 7\% in the range L9-T3. These numbers are qualitatively similar to \citet{Radigan2014}, who found a higher variability of 24$^{+11}_{-3}$\% for the L9-T3.5 range as compared to 3.2$^{+2.8}_{-1.8}$\% outside the L/T transition.

\subsection{Wide binary candidate systems}

We also search for binary systems constituted by two ultracool dwarfs (L+L,L+T,T+T). This type of system is very interesting, since widely separated ultracool dwarf binaries are quite rare, especially considering field ages. A large number of wide binary systems in the Galactic field could in fact rule out formation scenarios where very low-mass and substellar objects are ejected from the protocluster due to dynamical interactions \citep{Reipurth2001, Bate2005}. Due to their low binding energy, they are unlikely to survive this dynamical process.

A search for this type of binary system was previously presented in \citet{dalponte2020} using the sample of UCDs selected with the first DES data release. Here we used the same methodology and presented a new and updated list of this type of system. We used our UCD candidates catalog to search for binaries, where we computed a search radius for each UCD and checked if another ultracool dwarf appears inside this individual radius. The search radius was defined as a projected separation of 10,000 AU evaluated at the lower limit in distance of the UCD. 

For the initial list of candidates, we required that 
$\Delta_{\mu} \leq 2 \sigma_{\mu}$ where $\Delta_{\mu}$ is the total proper motion difference $$\Delta_{\mu} = \sqrt{\Delta_{\mu_{\alpha cos \delta}}^2 + \Delta_{\mu_{\delta}}^2}$$ and $\Delta_{\mu_{\alpha cos\delta}}$ and $\Delta_{\mu_{\delta}}$ are the differences in proper motion between the pair members. In the above criterion, $$\sigma_{\mu} = \sqrt{\delta \mu_{1}^2 + \delta \mu_{2}^2}$$ is the composite uncertainty in the measured proper motions, where 1,2 represent different objects of the system. We again required that each object has $\sigma_{\mu}/\mu$ $<$ 0.5. The next step was to demand common distances, using a criterion at the $3\sigma$ level. The final list has four new candidates and four already known, both with common proper motion and distance. Of these, we note that only one system (J030422-135839) has a common distance beyond $2\sigma$, which was the criterion adopted in \citet{dalponte2020}. However, this system has proper motions that are in clear agreement with each other. Table \ref{tab:comoving_systems} shows the new systems and those recovered from the literature. It is important to mention that not all systems presented in \citet{dalponte2020} were recovered here. The main reason is that some objects are now classified as M8 or M9 and therefore are not in the sample used for this new search.

To obtain the chance alignment probability we used the GalmodBD simulation code, which computes expected Galactic counts of ultracool dwarfs as a function of magnitude, colour and direction on the sky. The code also creates synthetic samples based on the expected number counts for a given footprint, using empirically determined space densities of objects, absolute magnitudes and colours as a function of spectral type. We computed the expected number of UCDs in a given direction and within the volume bracketed by the common range of distances and by the area within the angular separation of each system. For all the four new binary candidates, the probability of chance alignment is < 0.004\% .


\section{Spectroscopic confirmation of twelve ultracool dwarfs}
\label{sec:spectra}

We undertook a spectroscopic project to further assess our UCDs search and classification methods. We got $\sim$ 22 hours of Gemini/GMOS time to obtain spectra for a small fraction of our UCD candidates, twelve objects in total. The target sample for the spectrocopic follow up was taken from \citet{dalponte2020}. We have selected candidates that are more probable to be wide binary systems and for which the technical design will give us the best success rate. We also demanded the pair members to have a difference in distance modulus that was within 1.5 (1.0) from the typical expected difference given their uncertainties. Finally, we avoided the largest physical separation pairs to reach the final target sample. Our targets have magnitudes within the range 19 $<$ $z_{DES}$ $<$ 21.5 as shown in Table \ref{tab:ucd_properties_spectra} and the preference was given for the systems composed by two L dwarfs. The purpose of this follow up spectroscopy was to confirm their nature, i.e. confront spectral type with our photo-type method and also to re-calculate the distances.

\begin{table}
\centering
\caption{Observation log of the selected ultracool dwarfs. The central wavelength is in $\angstrom$ and the exposure time in minutes.}
\label{tab:obs_log}
\scalebox{0.7}{
\begin{tabular}[t]{lccccccl}
\hline \hline
Name & Obs. Date & Airmass & $\lambda$ (\angstrom) & Exp. Time \\
\hline
\multicolumn{5}{c}{GS-2019B-Q-230} \\
\hline
\multirow{4}{*}{UCD 10}  & 2019/09/30 & 1.35 & 7900 & 30.0 \\
    & 2019/10/07 & 1.37 & 7900 & 30.0 \\
    & 2019/10/07 & 1.28 & 8000 & 45.0 \\
    & 2019/10/07 & 1.22 & 8100 & 42.0 \\
    \\
\multirow{3}{*}{UCD 1}  & 2019/10/07 & 1.16 & 7900 & 30.0 \\
    & 2019/10/07 & 1.25 & 8000 & 68.8 \\
    & 2019/10/07 & 1.25 & 8100 & 42.0 \\
    \\
\multirow{3}{*}{UCD 3}  & 2019/11/30 & 1.06 & 7900 & 30.0 \\
    & 2019/11/30 & 1.07 & 8000 & 45.0 \\
    & 2019/11/30 & 1.12 & 8100 & 42.0 \\
    \\
\multirow{3}{*}{UCD 11} & 2019/12/01 & 1.36 & 7900 & 26.6 \\
    & 2019/12/01 & 1.50 & 8000 & 30.0 \\
    & 2019/12/01 & 1.73 & 8100 & 40.0 \\
\hline 
\multicolumn{5}{c}{GS-2019B-Q-230} \\
\hline
\multirow{3}{*}{UCD 36} & 2019/09/04 & 1.36 & 7900 & 60.0 \\
    & 2019/09/04 & 1.21 & 8000 & 60.0 \\
    & 2019/09/04 & 1.16 & 8100 & 45.0 \\
    \\
\multirow{3}{*}{UCD 8} & 2019/09/22  & 1.38 & 7900 & 30.0 \\
    & 2019/09/28 & 1.33 & 8000 & 30.0 \\
    & 2019/09/28 & 1.25 & 8100 & 30.0 \\
    \\
\multirow{4}{*}{UCD 6} & 2019/09/30  & 1.17 & 7900 & 60.0 \\
    & 2019/09/30 & 1.12 & 8000 & 22.16 \\
    & 2019/10/06 & 1.22 & 8000 & 30.0 \\
    & 2019/10/08 & 1.26 & 8100 & 45.0 \\
    \\
\multirow{4}{*}{UCD 12} & 2019/10/06  & 1.13 & 7900 & 60.0 \\
    & 2019/11/20 & 1.13 & 8000 & 15.0 \\
    & 2019/11/21 & 1.11 & 8000 & 45.0 \\
    & 2019/11/21 & 1.11 & 8100 & 45.0 \\
\hline \hline
\end{tabular}
}
\end{table}

\subsection{GMOS observation and data reduction}
\label{ref:gemini_ucd}

The selected UCDs were observed using the 8-m Gemini-South telescope with the Gemini Multi-Object Spectrograph \citep[GMOS, ][]{Hook2004}. The observations were carried out through the months of September to December 2019 as part of the programs GS-2019B-Q-230 (band 2) and GS-2019B-Q-312 (band 3).
We used GMOS with the R150 grating and the OG515 blocking filter to deliver a R$\sim$600 resolution, across the 7000-10000 $\angstrom$ range. For all targets, three spectra, centered at 7900, 8000, 8100 $\angstrom$ at z' filter were taken for each source, to cover the small gaps between the three GMOS detectors, and a focal plane unit of 1$\arcsec$ was selected. We binned both in spatial and spectral direction to 4x4 pixels to increase our signal-to-noise ratio. For every change in central wavelength, a flat and a CuAr frame was taken immediately following the science exposure. Table \ref{tab:obs_log} shows the observation log for all the objects observed with GMOS. The individual spectra for the same source were rebinned preserving flux and combined into a single coadded spectrum using standard routines. The typical signal-to-noise (SNR) per pixel for the coadd spectra is $\sim$ 6.

\begin{figure*}
\begin{center}
\includegraphics[width=0.9\linewidth]{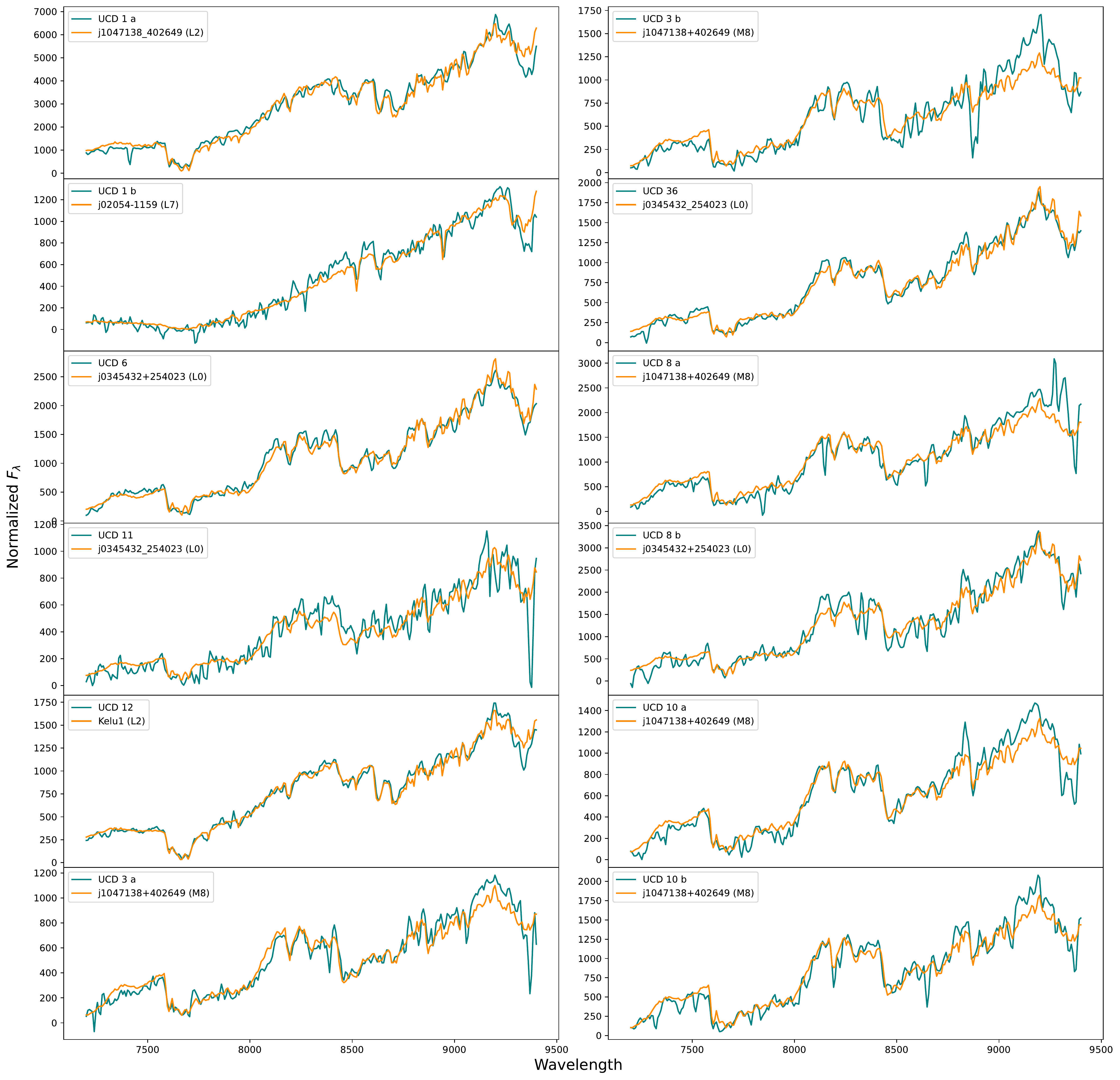}
\end{center}
\caption{UCD spectra (blue) and the best fitting template (orange), ordered by right ascension. The fluxes shown are relative $F_{\lambda}$ in arbitrary units. The flux of the templates was multiplied by a normalization factor prior to the fit, as explained previously. The individual members of the wide binary candidates are identified by \textit{a,b} labels.}
\label{fig:spectra_UCD_fit}
\end{figure*}

The objects UCD 1, UCD 3, UCD 10 and UCD 8 are wide binary systems candidates presented previously in \citet{dalponte2020} as composed by two L dwarfs. As an observation strategy, we place both objects (L dwarfs) of each system on a single long-slit to obtain two spectra at the same time. 
The data reduction was carried out using the standard GMOS pipeline contained in the GEMINI  \verb+IRAF/PyRAF+ package. The basic steps include bias subtraction, flat-field correction and wavelength calibration and for the extraction of the spectra we use the APALL pipeline. The spectra have not been flux calibrated and corrected for telluric absorption.

\subsection{Ultracool dwarfs spectral types}

To determine the spectral type for our UCDs we use a simple $\chi^2$ minimization relative to templates taken from \citet{Kirkpatrick1999}. The templates were smoothed down and rebinned to match our resolution and wavelength range of 7200-9400 $\angstrom$. We also visually inspected the five best-fitting templates to check the accuracy of the fit. For the instrumental fluxes, we attributed a Poisson fluctuation in the detector counts for every $\lambda$. We also multiplied the templates by a normalization factor before comparing them to each UCD spectrum. This normalization factor is given by

$$ N = \frac{\int_{\lambda_{1}}^{\lambda_{2}} Flux_{spectra}}{\int_{\lambda_{1}}^{\lambda_{2}} Flux_{template}} $$

\noindent where the integrals in the numerator and denominator are over the instrumental fluxes of GMOS and template spectra respectively, within the spectral range of our analysis ($\lambda=7200-9400$ $\angstrom$).

Figure \ref{fig:spectra_UCD_fit} shows the spectra and the lowest $\chi^2$ template, along with this best match spectral type. Table \ref{tab:ucd_properties_spectra} shows the photo-type estimated in \citet{carnero2019} and the new photo type estimation as presented earlier. The photo-type previously estimated has an typical uncertainty of one or two types due to the method adopted. 
As discussed previously, to obtain photometric distances we compared the photo-type with our empirical model grid to estimate the absolute magnitude and then obtained the distance modulus for each object. Now, with the use of the spectral type, new distances were estimated and are shown in the last column of Table.
UCD 1, UCD 3, and UCD 10 remain wide binary candidates based on the new distances measurements whereas UCD 8 is discarded as a common distance pair. This latter, in fact, is an interesting pair of sources. Their apparent magnitudes are quite similar in most filters and they are about 1 arcmin apart from each other on the sky. Their proper motion information comes from the CatWISE catalog and is not precise enough to help assessing the nature of the pair. On the other hand, their Gemini/GMOS spectra are best fit by an M8 and L0 template, respectively for the a and b components. In Appendix A we compare their spectra to other similar templates, showing the difficulty in attributing a spectral type with precision better than $\pm 2$ in some cases.

Also, the comoving candidate systems still have large uncertainties in their proper motion measurements, rendering current kinematical information not an efficient diagnostic. The spectra presented in this section are a basic sanity check that we are in fact selecting ultracool dwarfs and our method to estimate spectral types works as expected, with an accuracy of $\pm$ 2 types.

\begin{table*}
\centering
\caption{Objects z magnitude, photo-type, spectral type and photometric distances. The photo-type and the distance column is divided into \citet{carnero2019} and new measurements as presented in the earlier sections. All the estimates provided by \citet{carnero2019} have * mark.}
\label{tab:ucd_properties_spectra}
\scalebox{0.9}{
\begin{tabular}{lllccclll}
\hline \hline
Name & \multicolumn{1}{c}{RA} & \multicolumn{1}{c}{DEC} & \multicolumn{2}{c}{Ph.T} & Sp. Type & \multicolumn{1}{c}{z} & \multicolumn{2}{c}{Distance} \\
\hline
UCD 6 & 10.642 & -3.531 & L0* & L0 & L0 & 20.24$\pm$0.01 & 195 $\pm$ 33* & 176 $\pm$ 17  \\
UCD 11 & 21.102 & -44.986 & L2* & L1 & L0 & 20.92 $\pm$ 0.02 & 173 $\pm$ 8* & 219 $\pm$ 22 \\
UCD 12 & 22.886 & -5.240 & L2* & L1 & L2 & 20.83 $\pm$0.01 & 158 $\pm$ 9* & 149 $\pm$ 15 \\
UCD 36 & 321.070 & 0.484 & L0* & M9 & L0 & 20.63 $\pm$ 0.01 & 246 $\pm$ 10* & 222 $\pm$ 21 \\
\hline
UCD 1 a & 0.876 & -0.216 & L2* & L1 & L2 & 19.17$\pm$0.01  & 74 $\pm$ 2* & 70 $\pm$ 6 \\
UCD 1 b & 0.876 & -0.185 & L7* & L8 & L7 & 21.15$\pm$0.02 & 70 $\pm$ 7* & 64 $\pm$ 6 \\
UCD 3 a & 74.456 & -49.567 & L0* & M9 & M8 & 20.96 $\pm$ 0.02 & 349 $\pm$ 22* & 429 $\pm$ 50 \\
UCD 3 b & 74.455 & -49.565 & L0* & M9 & M8 & 21.13 $\pm$ 0.02 & 371 $\pm$ 29* & 457 $\pm$ 49 \\
UCD 8 a & 349.952 & -52.073 & L0* & M9 & M8 & 19.94 $\pm$ 0.01 & 181 $\pm$ 7* & 268 $\pm$ 26 \\
UCD 8 b & 349.929 & -52.065 & L0* & M9 & L0 & 19.95 $\pm$ 0.01 & 176 $\pm$ 5* & 160 $\pm$ 15 \\
UCD 10 a & 355.516 & -61.588 & L0* & M9 & M8 & 20.78 $\pm$ 0.01 & 279 $\pm$ 29* & 416 $\pm$ 41 \\
UCD 10 b & 355.533 & -61.595 & L0* & L0 & M8 & 21.17 $\pm$ 0.02 & 286 $\pm$ 29* & 445 $\pm$ 48 \\
\hline \hline
\end{tabular}
}
\end{table*}

\section{Conclusions}
\label{sec:conclusions}

Using the recent Dark Energy Survey data release (DR2) combined with VHS DR6, VIKING DR5 and AllWISE data, we were able to identify new ultracool dwarfs candidates, probing faint and more distant objects than those presented in the literature so far. We select these candidates based on their colors ($i$-$z$), ($z$-$Y$), and ($Y$-$J$) up to $z$ $\leq$ 23. We applied a classification method where a photo-type can be attributed to each object based only on its photometry. Here we have presented updated colour templates in our classification scheme compared to previous work in \citet{carnero2019}, and expanded the ultracool dwarf candidate sample to cover almost the entire DES footprint area, thanks to the new VHS DR6 catalog.

In total, our new sample has 19,583 ultracool dwarf candidates, where 14,099 are presented here for the first time. The complete sample includes 142 spectroscopically confirmed objects from the literature, plus 5,342 ultracool dwarf candidates from the literature, where the vast majority (5,257 candidates) are from our previous work. The samples from the literature, both with spectroscopic confirmation and candidates were used here as a validation to our method. The method to infer the spectral type consists in a minimization of the $\chi^{2}$ relative to empirical templates of M, L and T dwarfs. The comparison between our estimated photo-type with those from the literature showed us that our photo-type is accurate in $\pm$2 spectral types. During the classification step, we also used Lephare code with templates from galaxies and quasars in order to identify extragalactic contamination and remove those sources from our final sample.

Our L and T candidates comprise the largest such sample as of today. For instance, \citet{Skrzypek2016} report on finding 1361 L and T dwarfs brighter than J = 17.5 within an effective area of 3070 deg2 in the Northern Hemisphere. Their UCDs span distances out to ~150 pc, whereas our sample goes at least 3 times farther, out to $\sim$ 500pc, and covers a solid angle 60\% larger. This much larger volume, coupled with the exponential drop in density in the Galactic disk at the high latitudes we cover, make the two samples quite consistent in terms of the number of objects found.

We also show some applications for our new ultracool dwarf candidates: i) search for new candidate members to young moving groups and associations; ii) photometric variable objects; iii) search for new wide binary candidate systems. For the first application, we used the BANYAN $\Sigma$ algorithm to investigate the likelihood of each object in our sample being a member of a young moving group. We found 20 new candidate members with membership probability > 90\%. We also identify 291 variable candidate sources in our sample, of which 10\% are young objects. Also, a higher percentage of the variable sample is concentrated in the L9-T3 range. In addition, as presented previously in \citet{dalponte2020}, we search for wide binary systems composed of two ultracool dwarfs and here we present four new candidate systems.

Finally, we show here the spectroscopic confirmation of twelve new ultracool dwarfs, a basic sanity check of our selection and classification method. 

\section{Acknowledgments}

Funding for the DES Projects has been provided by the U.S. Department of Energy, the U.S. National Science Foundation, the Ministry of Science and Education of Spain, 
the Science and Technology Facilities Council of the United Kingdom, the Higher Education Funding Council for England, the National Center for Supercomputing 
Applications at the University of Illinois at Urbana-Champaign, the Kavli Institute of Cosmological Physics at the University of Chicago, the Center for Cosmology and Astro-Particle Physics at the Ohio State University, the Mitchell Institute for Fundamental Physics and Astronomy at Texas A\&M University, Financiadora de Estudos e Projetos, 
Funda{\c c}{\~a}o Carlos Chagas Filho de Amparo {\`a} Pesquisa do Estado do Rio de Janeiro, Conselho Nacional de Desenvolvimento Cient{\'i}fico e Tecnol{\'o}gico and the Minist{\'e}rio da Ci{\^e}ncia, Tecnologia e Inova{\c c}{\~a}o, the Deutsche Forschungsgemeinschaft and the Collaborating Institutions in the Dark Energy Survey. 

The Collaborating Institutions are Argonne National Laboratory, the University of California at Santa Cruz, the University of Cambridge, Centro de Investigaciones Energ{\'e}ticas, 
Medioambientales y Tecnol{\'o}gicas-Madrid, the University of Chicago, University College London, the DES-Brazil Consortium, the University of Edinburgh, the Eidgen{\"o}ssische Technische Hochschule (ETH) Z{\"u}rich, Fermi National Accelerator Laboratory, the University of Illinois at Urbana-Champaign, the Institut de Ci{\`e}ncies de l'Espai (IEEC/CSIC), 
the Institut de F{\'i}sica d'Altes Energies, Lawrence Berkeley National Laboratory, the Ludwig-Maximilians Universit{\"a}t M{\"u}nchen and the associated Excellence Cluster Universe, the University of Michigan, the NSF's National Optical-Infrared Astronomy Research Laboratory, the University of Nottingham, The Ohio State University, the University of Pennsylvania, the University of Portsmouth, SLAC National Accelerator Laboratory, Stanford University, the University of Sussex, Texas A\&M University, and the OzDES Membership Consortium.

Based in part on observations at Cerro Tololo Inter-American Observatory, NSF's National Optical-Infrared Astronomy Research Laboratory, which is operated by the Association of Universities for Research in Astronomy (AURA) under a cooperative agreement with the National Science Foundation.

The DES data management system is supported by the National Science Foundation under Grant Numbers AST-1138766 and AST-1536171. The DES participants from Spanish institutions are partially supported by MINECO under grants AYA2015-71825, ESP2015-66861, FPA2015-68048, SEV-2016-0588, SEV-2016-0597, and MDM-2015-0509, some of which include ERDF funds from the European Union. IFAE is partially funded by the CERCA program of the Generalitat de Catalunya. Research leading to these results has received funding from the European Research
Council under the European Union's Seventh Framework Program (FP7/2007-2013) including ERC grant agreements 240672, 291329, and 306478. We  acknowledge support from the Australian Research Council Centre of Excellence for All-sky Astrophysics (CAASTRO), through project number CE110001020, and the Brazilian Instituto Nacional de Ci\^encia e Tecnologia (INCT) e-Universe (CNPq grant 465376/2014-2).

This manuscript has been authored by Fermi Research Alliance, LLC under Contract No. DE-AC02-07CH11359 with the U.S. Department of Energy, Office of Science, Office of High Energy Physics. The United States Government retains and the publisher, by accepting the article for publication, acknowledges that the United States Government retains a non-exclusive, paid-up, irrevocable, world-wide license to publish or reproduce the published form of this manuscript, or allow others to do so, for United States Government purposes.

This publication makes use of data products from the Wide-field Infrared Survey Explorer, which is a joint project of the University of California, Los Angeles, and the Jet Propulsion Laboratory/California Institute of Technology, and NEOWISE, which is a project of the Jet Propulsion Laboratory/California Institute of Technology. WISE and NEOWISE are funded by the National Aeronautics and Space Administration. 

The analysis presented here is based on observations obtained as part of the VISTA Hemisphere Survey, ESO Programme, 179.A-2010 (PI: McMahon). 

This paper has gone through internal review by the DES collaboration.

MDP acknowledges financial support provided by the CNPq Fellowship.

\section{Data availability}

Data underlying this article are available in \url{https://des.ncsa.illinois.edu/releases/other/Y6-LTdwarfs}.

\bibliography{references}

\section{Affiliations}

$^{1}$ Instituto de F\'\i sica, UFRGS, Caixa Postal 15051, Porto Alegre, RS - 91501-970, Brazil \\
$^{2}$ Laborat\'orio Interinstitucional de e-Astronomia - LIneA, Rua Gal. Jos\'e Cristino 77, Rio de Janeiro, RJ - 20921-400, Brazil \\
$^{3}$ Instituto de Astrofisica de Canarias, E-38205 La Laguna, Tenerife, Spain \\
$^{4}$ Universidad de La Laguna, Dpto. Astrofísica, E-38206 La Laguna, Tenerife, Spain \\
$^{5}$ McWilliams Center for Cosmology, Carnegie Mellon University, 5000 Forbes Ave, Pittsburgh, PA 15213, USA \\
$^{6}$ Physics Department, University of Wisconsin-Madison, Madison, WI 53706, USA \\
$^{7}$ Cerro Tololo Inter-American Observatory, NSF's National Optical-Infrared Astronomy Research Laboratory, Casilla 603, La Serena, Chile \\
$^{8}$ Fermi National Accelerator Laboratory, P. O. Box 500, Batavia, IL 60510, USA \\
$^{9}$ Department of Physics, University of Michigan, Ann Arbor, MI 48109, USA \\
$^{10}$ Institute of Cosmology and Gravitation, University of Portsmouth, Portsmouth, PO1 3FX, UK \\
$^{11}$ CNRS, UMR 7095, Institut d'Astrophysique de Paris, F-75014, Paris, France \\
$^{12}$ Sorbonne Universit\'es, UPMC Univ Paris 06, UMR 7095, Institut d'Astrophysique de Paris, F-75014, Paris, France \\
$^{13}$ University Observatory, Faculty of Physics, Ludwig-Maximilians-Universit\""at, Scheinerstr. 1, 81679 Munich, Germany \\
$^{14}$ Department of Physics \& Astronomy, University College London, Gower Street, London, WC1E 6BT, UK \\
$^{15}$ Kavli Institute for Particle Astrophysics \& Cosmology, P. O. Box 2450, Stanford University, Stanford, CA 94305, USA \\
$^{16}$ SLAC National Accelerator Laboratory, Menlo Park, CA 94025, USA \\
$^{17}$ Center for Astrophysical Surveys, National Center for Supercomputing Applications, 1205 West Clark St., Urbana, IL 61801, USA \\
$^{18}$ Department of Astronomy, University of Illinois at Urbana-Champaign, 1002 W. Green Street, Urbana, IL 61801, USA \\
$^{19}$ Institut de F\'{\i}sica d'Altes Energies (IFAE), The Barcelona Institute of Science and Technology, Campus UAB, 08193 Bellaterra (Barcelona) Spain \\
$^{20}$ Jodrell Bank Center for Astrophysics, School of Physics and Astronomy, University of Manchester, Oxford Road, Manchester, M13 9PL, UK \\
$^{21}$ University of Nottingham, School of Physics and Astronomy, Nottingham NG7 2RD, UK \\
$^{22}$ Astronomy Unit, Department of Physics, University of Trieste, via Tiepolo 11, I-34131 Trieste, Italy \\ 
$^{23}$ INAF-Osservatorio Astronomico di Trieste, via G. B. Tiepolo 11, I-34143 Trieste, Italy \\
$^{24}$ Institute for Fundamental Physics of the Universe, Via Beirut 2, 34014 Trieste, Italy \\
$^{25}$ Department of Physics, IIT Hyderabad, Kandi, Telangana 502285, India \\
$^{26}$ Centro de Investigaciones Energ\'eticas, Medioambientales y Tecnol\'ogicas (CIEMAT), Madrid, Spain \\
$^{27}$ Jet Propulsion Laboratory, California Institute of Technology, 4800 Oak Grove Dr., Pasadena, CA 91109, USA \\
$^{28}$ Institute of Theoretical Astrophysics, University of Oslo. P.O. Box 1029 Blindern, NO-0315 Oslo, Norway \\
$^{29}$ Kavli Institute for Cosmological Physics, University of Chicago, Chicago, IL 60637, USA \\
$^{30}$ Instituto de Fisica Teorica UAM/CSIC, Universidad Autonoma de Madrid, 28049 Madrid, Spain \\
$^{31}$ Department of Astronomy, University of Michigan, Ann Arbor, MI 48109, USA \\
$^{32}$ School of Mathematics and Physics, University of Queensland,  Brisbane, QLD 4072, Australia \\
$^{33}$ Santa Cruz Institute for Particle Physics, Santa Cruz, CA 95064, USA \\ 
$^{34}$ Center for Astrophysics $\vert$ Harvard \& Smithsonian, 60 Garden Street, Cambridge, MA 02138, USA \\
$^{35}$ Australian Astronomical Optics, Macquarie University, North Ryde, NSW 2113, Australia \\
$^{36}$ Lowell Observatory, 1400 Mars Hill Rd, Flagstaff, AZ 86001, USA \\
$^{37}$ George P. and Cynthia Woods Mitchell Institute for Fundamental Physics and Astronomy, and Department of Physics and Astronomy, Texas A\&M University, College Station, TX 77843,  USA \\
$^{38}$ Instituci\'o Catalana de Recerca i Estudis Avan\c{c}ats, E-08010 Barcelona, Spain \\
$^{39}$ Observat\'orio Nacional, Rua Gal. Jos\'e Cristino 77, Rio de Janeiro, RJ - 20921-400, Brazil \\
$^{40}$ Department of Astronomy, University of California, Berkeley,  501 Campbell Hall, Berkeley, CA 94720, USA \\
$^{41}$ Institute of Astronomy, University of Cambridge, Madingley Road, Cambridge CB3 0HA, UK \\
$^{42}$ Hamburger Sternwarte, Universit\""{a}t Hamburg, Gojenbergsweg 112, 21029 Hamburg, Germany \\
$^{43}$ Department of Astrophysical Sciences, Princeton University, Peyton Hall, Princeton, NJ 08544, USA \\
$^{44}$ Department of Physics and Astronomy, University of Pennsylvania, Philadelphia, PA 19104, USA \\
$^{45}$ School of Physics and Astronomy, University of Southampton,  Southampton, SO17 1BJ, UK \\
$^{46}$ Computer Science and Mathematics Division, Oak Ridge National Laboratory, Oak Ridge, TN 37831 \\
$^{47}$ Center for Cosmology and Astro-Particle Physics, The Ohio State University, Columbus, OH 43210, USA \\
$^{48}$ Lawrence Berkeley National Laboratory, 1 Cyclotron Road, Berkeley, CA 94720, USA \\

\appendix

\section{UCD 8 objects template fitting}
\label{app:spectra_templates_ucd8}

\begin{figure*}
\centering
\includegraphics[width=0.9\linewidth]{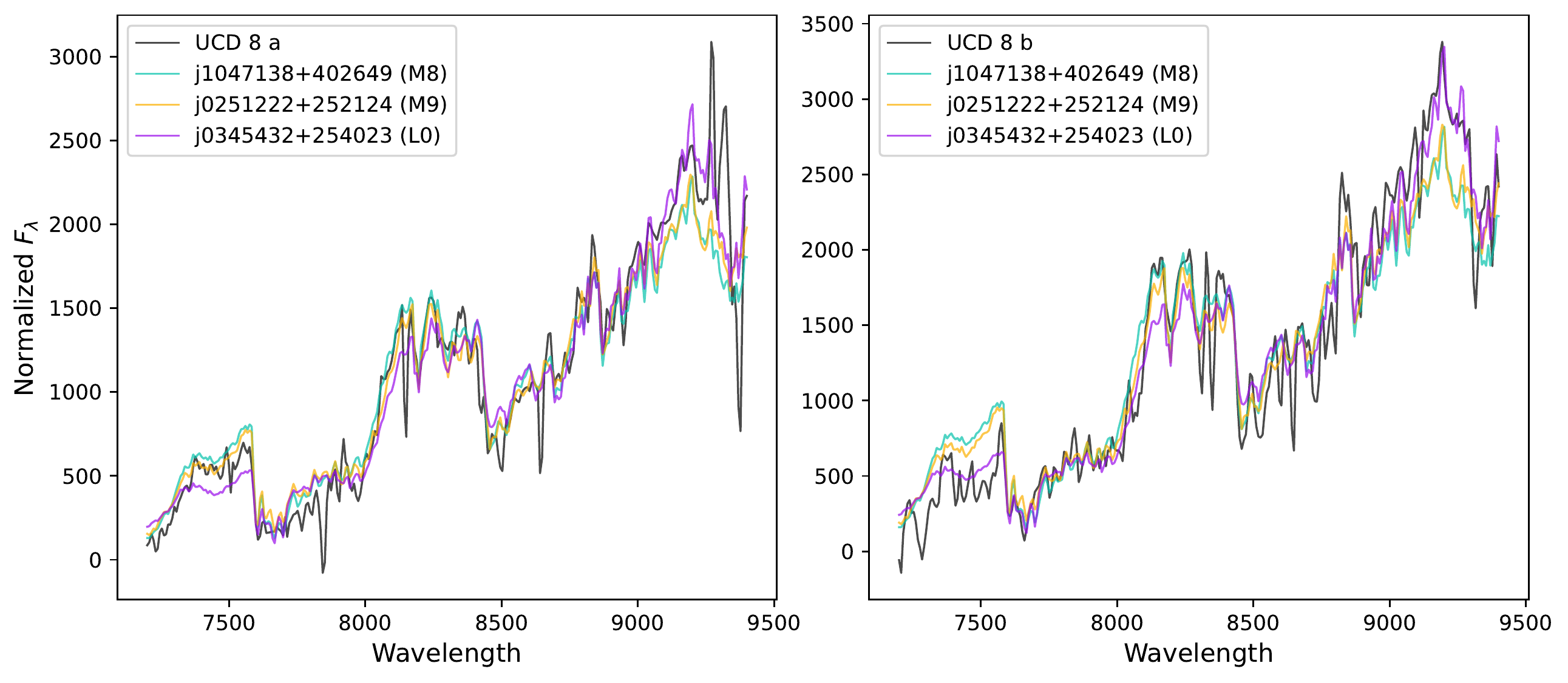}
\caption{UCD spectra (black) and the M8 (green), M9 (yellow) and L0 (purple) templates. The fluxes shown are relative $F_{\lambda}$ in arbitrary units. As in Figure \ref{fig:spectra_UCD_fit}, the flux of the templates was multiplied by a normalization factor.}
\label{fig:ucd8_templates}
\end{figure*}

In assessing the nature of binary system candidates involving one or more UCD, it is important to take into account the uncertainties in assigning a spectral type, since our photometric distances are based on template absolute magnitudes for each type. We show here the interesting example of the pair made by UCD 8, for which we obtained Gemini/GMOS spectra of the UCD 8a and UCD 8b candidate members. Figure \ref{fig:ucd8_templates} we show these spectra along with the templates for M8, M9 and L0. The M8 and M9 templates fit the spectra in a very similar way. The L0 template seems to be the best fit for the UCD 8 b, while for the UCD 8 a, the M8 has the lowest $\chi^2$. As mentioned in Section \ref{sec:spectra}, these two sources are particularly interesting since the objects have very similar magnitudes in all bands and we could expect a more similar spectral type. Because the assigned spectral types differ by two units, however, their distances are now inconsistent with a physical pair. More accurate distance estimates, as well as proper motions, will be required to resolve their nature.

\end{document}